\documentclass[journal]{IEEEtran}

\usepackage[utf8]{inputenc}
\usepackage[T1]{fontenc}
\usepackage{url}
\usepackage{ifthen}
\usepackage{cite}
\usepackage[cmex10]{amsmath}

\usepackage{graphicx}
\usepackage{amsfonts}
\usepackage{amsmath}
\usepackage{amssymb}
\usepackage{mathrsfs} 
\usepackage{color}
\usepackage{bm}
\usepackage{latexsym} 

\newtheorem{thm}{Theorem}

\newtheorem{lem}{Lemma}
\newtheorem{proof}{proof}

\newtheorem{defn}{Definition}
\newtheorem{rem}{Remark}
\newtheorem{exam}{Example}

\ifCLASSINFOpdf \else \fi \onecolumn 

\begin{document}
\title{Sequence Reconstruction for the Single-Deletion
Single-Substitution Channel}

\author{Wentu~Song, ~Kui~Cai,~\IEEEmembership{Senior
        Member,~IEEE},
        and Tony Q. S. Quek,~\IEEEmembership{Fellow,~IEEE}
\thanks{Wentu~Song, Kui~Cai and Tony Q. S. Quek are with Singapore
        University of Technology and Design, Singapore, e-mail:
        \{wentu\_song, cai\_kui, tonyquek\}@sutd.edu.sg.
        (\emph{Corresponding author: Kui Cai.)}}
}

\maketitle

\begin{abstract}
The central problem in sequence reconstruction is to find the
minimum number of distinct channel outputs required to uniquely
reconstruct the transmitted sequence. According to Levenshtein's
work in 2001, this number is determined by the size of the maximum
intersection between the error balls of any two distinct input
sequences of the channel. In this work, we study the sequence
reconstruction problem for single-deletion single-substitution
channel, assuming that the transmitted sequence belongs to a
$q$-ary code with minimum Hamming distance at least $2$, where
$q\geq 2$ is any fixed integer. Specifically, we prove that for
any two $q$-ary sequences of length $n$ and with Hamming distance
$d\geq 2$, the size of the intersection of their error balls is
upper bounded by $2qn-3q-2-\delta_{q,2}$, where $\delta_{i,j}$ is
the Kronecker delta. We also prove the tightness of this bound by
constructing two sequences the intersection size of whose error
balls achieves this bound.
\end{abstract}


\IEEEpeerreviewmaketitle

\section{Introduction}

\IEEEPARstart{W}{e} consider a communication scenario where a
codeword $\bm x$ from some codebook $\mathscr C$ is transmitted
over a number of identical channels and the goal is to reconstruct
$\bm x$ from all (erroneous) channel outputs (also referred to as
reads in data storage applications). This problem, also known as
the sequence reconstruction problem, was first proposed by
Levenshtein \cite{Levenshtein01-IT}, \cite{Levenshtein01-CTA}, and
in recent years, gained more and more attentions due to its
applications in DNA data storage \cite{Sabary24-MBC}. The central
problem in sequence reconstruction is to determine the minimum
number of distinct channel outputs (reads) required to uniquely
reconstruct $\bm x$. This number was shown to be equal to one plus
the size of the maximum intersection between the error balls of
any two distinct codewords of $\mathscr C$ (also referred to as
the read coverage of $\mathscr C$ for the corresponding channel)
\cite{Levenshtein01-IT}. Therefore, deriving the read coverage of
$\mathscr C$ is critical to solving the sequence reconstruction
problem. On the other hand, designing codes with given read
coverage, called reconstruction codes, is also an interesting
problem for sequence reconstruction.

In his seminal work \cite{Levenshtein01-IT}, Levenshtein studied
the sequence reconstruction problem for deletion, insertion,
substitution and transposition separately, where $\mathscr C$ is
taken to be the set of all $q$-ary sequence. For the more general
case that $\mathscr C$ is an $(\ell-1)$-deletion correcting code
for some positive integer $\ell\leq t$, the problem was studied in
\cite{Gabrys18-SR} and \cite{Pham22} for $t$-deletion channel, and
in \cite{Sala17-IT} for $t$-insertion channel. Reconstruction
codes for two-deletion channels can be found in
\cite{Chrisnata22-IT, Y-Sun-23-SRCC} and reconstruction codes for
two-insertion channels can be found in \cite{Z-L-G-23}.
Reconstruction codes for $q$-ary single-edit channel $(q\geq 2)$
was constructed in \cite{KuiCai22} by generalizing the
construction in \cite{Chee18}, where an edit error means a
deletion, an insertion or a substitution error. Reconstruction
codes for single-burst-insertion/deletion were constructed in
\cite{Y-Sun-23}, where a burst of $t$ deletions/insertions means
$t$ deletions or $t$ insertions occurring at consecutive
positions. In these constructions, each read is corrupted by only
one type of error.

In practical applications, a read may suffers from different error
types, for example, both a deletion and an insertion, or both a
deletion and a substitution. It was shown in \cite{Levenshtein65}
that a code $\mathscr C$ can correct $t$ deletions if and only if
it can correct $t$ insertions. However, the intersection size of
$t$-deletion balls of two sequences is not necessarily equal to
the intersection size of their $t$-insertion balls when the
intersections are not empty. Therefore, unlike the classic error
correction problem, in sequence reconstruction problem, the
deletion channel and the insertion channel must be treated
separately. The reconstruction problem for single-insertion
single-substitution was studied in \cite{Abu-Sini21}, where the
maximum intersection size of binary single-insertion
single-substitution balls was proved to be
$\lfloor\frac{n-2}{2}\rfloor\lceil\frac{n-2}{2}\rceil+4n$. The
size of single-deletion multiple-substitution ball was also
computed in \cite{Abu-Sini21}, but their intersection size was not
considered. In a more recent work \cite{Abbasian24}, the size of
the error ball for $q$-ary channels with multiple types of errors
and at most three edits was studied. To the best of our knowledge,
deriving the maximum intersection size of single-deletion
single-substitution balls is still an open problem.

In this work, we study the sequence reconstruction problem for
$q$-ary single-deletion single-substitution channel, where $q\geq
2$ is an arbitrarily fixed integer. For example, in DNA data
storage, $q$ is usually taken to be $4$. We prove that for any two
$q$-ary sequences with Hamming distance $d\geq 2$, the size of the
intersection of their error balls is upper bounded by
$2qn-3q-2-\delta_{q,2}$, where $\delta_{i,j}$ is the Kronecker
delta. We also show that there exist two sequences the
intersection size of whose error balls achieves this bound, which
proves that the bound is tight. Note that the requirement that the
Hamming distance between two sequences is at least $2$ can be
satisfied by adding one parity check symbol, so results in one
symbol of redundancy.

The single-deletion single-substitution channel has been studied
in several existing works under the classic error correction model
or list-decoding model \cite{Smagloy24-IT}$-$\cite{Sun-Ge24}. The
best known single-deletion single-substitution correcting binary
codes has $4\log n+o(\log n)$ bits of redundancy, where $n$ is the
code length \cite{Sun-Ge24}. By our result, when the number of
reads is $2qn-3q-1-\delta_{q,2}$, one symbol of redundancy is
sufficient to guarantee correct reconstruction of the transmitted
sequence.

The paper is organized as follows. In Section
\uppercase\expandafter{\romannumeral 2}, we describe the problem
and our main result, as well as some simple observations that will
help to prove our main result. In Section
\uppercase\expandafter{\romannumeral 3}, we develop a method for
dividing the intersection of two error balls into some subsets
whose size can be easily obtained. We give a formal proof of our
main result in Section \uppercase\expandafter{\romannumeral 4} and
make conclusions and some discussions for future work in Section
\uppercase\expandafter{\romannumeral 5}.

\section{Problem description and main result}

For any integers $m\leq n$, let $[m,n]=\{m,m+1,\ldots,n\}~($called
an \emph{interval}$)$ and let $[n]=[1,n]$. For any set $A$, $|A|$
is the size of $A$; if $A$ is a set of numbers, then
$\min(A)~($resp. $\max(A))$ is the smallest $($resp. greatest$)$
number in $A$. For simplicity, we denote $A\backslash
i=A\backslash\{i\}$ for any $i\in A$. If we denote
$A=\{i_1,i_2,\cdots,i_k\}$, we always assume that
$i_1<i_2<\cdots<i_k$.

Let $\Sigma_q=\{0,1,\cdots,q-1\}$, where $q\geq 2$ is an
arbitrarily fixed integer. For any $\bm x\in\Sigma_q^n$, let $x_i$
denote the $i$th component of $\bm x$ and write $\bm
x=x_1x_2\cdots x_n$ or $\bm x=(x_1,x_2,\cdots, x_n)$. If
$D=\{i_1,i_2,\cdots,i_m\}\subseteq[n]$, let
$x_D=x_{i_1}x_{i_2}\cdots x_{i_m}$ and call it a
\emph{subsequence} of $\bm x$. If $D$ is an interval, $x_D$ is
called a \emph{substring} of $\bm x$. A \emph{run} of $\bm x$ is a
maximal substring of $\bm x$ consisting of identical symbols. For
any two given distinct symbols $a,b\in\Sigma_q$, let $A_{n}(ab)$
denote the \emph{alternating sequence} of length $n$ that starts
with $a$ and consists of $a,b$. For example, $A_5(ab)=ababa$ and
$A_6(ab)=ababab$.

For any $\bm x,\bm x'\in\Sigma_q^n$, the \emph{Hamming distance}
between $\bm x$ and $\bm x'$, denoted by $d_{\text{H}}(\bm x,\bm
x')$, is defined as the number of $i\in[n]$ such that $x_i\neq
x'_i$. The \emph{Levenshtein distance} between $\bm x$ and $\bm
x'$, denoted by $d_{\text{L}}(\bm x,\bm x')$, is defined as the
smallest integer $\ell$ such that $\bm x$ and $\bm x'$ share some
subsequence of length $n-\ell$.

Let $t$ and $s$ be non-negative integers such that $t+s<n$. For
any $\bm{x}\in\Sigma_q^n$, the \emph{$t$-deletion $s$-substitution
ball} of $\bm x$, denoted by $B^{\text{D,S}}_{t,s}(\bm{x})$, is
the set of all sequences that can be obtained from $\bm{x}$ by
exact $t$ deletions and at most $s$ substitutions. The
\emph{$t$-deletion ball} of $\bm x$ is
$B^{\text{D}}_{t}(\bm{x})\triangleq B^{\text{D,S}}_{t,0}(\bm{x})$,
and the \emph{$s$-substitution ball} of $\bm x$ is
$B^{\text{S}}_{s}(\bm{x})\triangleq B^{\text{D,S}}_{0,s}(\bm{x})$.
For $B\in\{B^{\text{D}}_{t}, B^{\text{S}}_{s},
B^{\text{D,S}}_{t,s}\}$, let $B(\bm x,\bm x')\triangleq B(\bm x)
\cap B(\bm x').$ Given a code $\mathscr C\subseteq\Sigma_q^n$, let
$$\nu(\mathscr C; B)\triangleq\max\{|B(\bm x,\bm x')|:
\bm x,\bm x'\in\mathscr C,\bm x\neq\bm x'\}$$ called the
\emph{read coverage} of $\mathscr C$ with respect to $B$. A
central problem in sequence reconstruction is to compute
$\nu(\mathscr C; B)$, given $\mathscr C$ and $B$. Another problem
is, given the error ball function $B$ and a positive integer $N$,
to design a code $\mathscr C\subseteq\Sigma_q^n$ with
$\nu(\mathscr C;B)<N$, called an $(n,N,B)$-reconstruction code.

In this work, we assume $q\geq 2$ is any fixed positive integer
and consider the sequence reconstruction problem for $q$-ary
single-deletion single-substitution channel $($i.e.,
$B=B^{\text{D,S}}_{1,1})$. Our main result is the following
theorem.

\begin{thm}\label{thm-ins-size}
Suppose $n\geq\max\{\frac{q+23}{2}, \frac{5q+19}{q-1}\}$. For any
$\bm x,\bm x'\in\Sigma_q^n$ with $d_{\text{H}}(\bm x,\bm x')\geq
2$, we have
$$|B^{\text{D,S}}_{1,1}(\bm x,\bm x')|\leq 2qn-3q-2-\delta_{q,2}$$
where $\delta_{i,j}$ is the Kronecker delta. Moreover, there exist
two sequences $\bm x,\bm x'\in\Sigma_q^n$ with $d_{\text{H}}(\bm
x,\bm x')=2$ and $|B^{\text{D,S}}_{1,1}(\bm x,\bm
x')|=2qn-3q-2-\delta_{q,2}$.
\end{thm}

The proof of Theorem \ref{thm-ins-size} will be given in Section
IV. In the rest of this section, we state some simple observations
that will be used in our proof.



\subsection{Intersection size of error balls of $q$-ary
substitution channel}

First, the size of the $q$-ary substitution ball satisfies
$($e.g., see \cite[Chapter 1]{Huffman}$)$
$$|B^{\text{S}}_{s}(\bm x)|=\sum_{k=0}^{s}\binom{n}{k}(q-1)^k,
~\forall\!~\bm x\in\Sigma_q^n.$$ In particular, for $s=1$, we have
\begin{equation}\label{eq-size-BS1}
|B^{\text{S}}_{1}(\bm x)|=1+(q-1)n,~\forall\!~\bm
x\in\Sigma_q^n.\end{equation} For the intersection size of
single-substitution balls, we have the following simple remark.
\begin{rem}\label{rem-sub-int-size}
For any $\bm x,\bm x'\in\Sigma_q^n$, we have
\begin{equation*}
|B^{\text{S}}_{1}(\bm x,\bm x')|=\!\left\{\!\begin{aligned}
&q, ~\text{if}~d_{\text{H}}(\bm x,\bm x')=1;\\
&2, ~\text{if}~d_{\text{H}}(\bm x,\bm x')=2;\\
&0, ~\text{if}~d_{\text{H}}(\bm x,\bm x')\geq 3.
\end{aligned}\right.
\end{equation*}
\end{rem}

\subsection{Some useful observations and lemma}

Consider the intersection size of $t$-deletion $s$-substitution
balls. Suppose $\bm x,\bm x'\in\Sigma_q^n$. By the definition of
$B^{\text{D,S}}_{t,s}$, it is easy to see that
$$B^{\text{D,S}}_{t,s}(\bm x,\bm x')
=\bigcup_{\bm z\in B^{\text{D}}_{t}(\bm x), \bm z'\in
B^{\text{D}}_{t}(\bm x')}B^{\text{S}}_{s}(\bm z,\bm z').$$ For the
special case that $t=s=1$, we have
\begin{align*}B^{\text{D,S}}_{1,1}(\bm x,\bm x')=\bigcup_{\bm z\in
B^{\text{D}}_{1}(\bm x), \bm z'\in B^{\text{D}}_{1}(\bm
x')}B^{\text{S}}_{1}(\bm z,\bm
z')=\bigcup_{j,j'\in[n]}B^{\text{S}}_{1}(x_{[n]\backslash
j},x'_{[n]\backslash j'}).\end{align*} Note that by Remark
\ref{rem-sub-int-size}, $|B^{\text{S}}_{1}(\bm z,\bm z')|=0$ when
$d_{\text{H}}(\bm z,\bm z')\geq 3$. Then we have the following
observation:

\textbf{Observation 1}: It holds that
$$B^{\text{D,S}}_{1,1}(\bm x,\bm x')=\bigcup_{(\bm z,\bm
z')\in\Lambda}B^{\text{S}}_{1}(\bm z,\bm z')$$ where
\begin{align}\label{eq-def-LMD}
\Lambda=\Lambda(\bm x,\bm x')\triangleq\Big\{(x_{[n]\backslash
j},x'_{[n]\backslash j'}): j,j'\in[n] ~\text{and}~
d_{\text{H}}(x_{[n]\backslash j}, x'_{[n]\backslash j'})\leq
2\Big\}.\end{align}

To compute $d_{\text{H}}(x_{[n]\backslash j},x'_{[n]\backslash
j'})$, we introduce some notations as follows. Let
\begin{align}\label{eq-def-S} S=S(\bm x,\bm
x')\triangleq\{i\in[n]: x_i\neq x'_{i}\}.\end{align} Then we have
$|S|=d_{\text{H}}(\bm x, \bm x')$, and so we can denote
$$S=\{i_1,i_2,\cdots,i_d\}$$ where $d=d_{\text{H}}(\bm x, \bm x')$
and $i_1<i_2<\cdots<i_d$ according to our previous convention. We
further let
\begin{align}\label{eq-def-TL}
T^L=T^L(\bm x,\bm x')\triangleq\{i\in[2,n]: x_i\neq
x'_{i-1}\}\end{align} and
\begin{align}\label{eq-def-TR}T^R=T^R(\bm x,\bm
x')\triangleq\{i\in[2,n]: x_{i-1}\neq x'_{i}\}.\end{align} From
these definitions, it is easy to see that $T^R(\bm x,\bm
x')=T^L(\bm x',\bm x)$. Note that in the notations $T^L(\bm x,\bm
x')$ and $T^R(\bm x,\bm x')$, $(\bm x,\bm x')$ is viewed as an
ordered pair. Moreover, by the definitions, $T^L=T^L(\bm x,\bm
x')\neq T^L(\bm x',\bm x)$ and $T^R=T^R(\bm x,\bm x')\neq T^R(\bm
x',\bm x)$.

Now, we have the second useful observation.

\textbf{Observation 2}: For any $j,j'\in[n]$, $j\leq j'$, we have
\begin{align*}d_{\text{H}}(x_{[n]\backslash j},x'_{[n]\backslash
j'})&=\left|\big(S\cap[1,j-1]\big)\cup
\big(T^L\cap[j+1,j']\big)\cup\big(S\cap[j'+1,n]\big)\right|
\\&=|S\cap[1,j-1]|+|T^L\cap[j+1,j']|+|S\cap[j'+1,n]|;
\end{align*} and
\begin{align*}d_{\text{H}}(x_{[n]\backslash j'},x'_{[n]\backslash
j})&=\left|\big(S\cap[1,j-1]\big)\cup
\big(T^R\cap[j+1,j']\big)\cup\big(S\cap[j'+1,n]\big)\right|
\\&=|S\cap[1,j-1]|+|T^R\cap[j+1,j']|+|S\cap[j'+1,n]|.
\end{align*}

The following lemma will be used to exclude repeat count of
sequence pairs in $\Lambda$.

\begin{lem}\label{iden-eq-dH}
Suppose $j_1,j_2,j_1',j_2'\in[n]$ such that $j_1\leq j_2$ and
$j_1'\leq j_2'$. The following hold.
\begin{itemize}
 \item[1)] If $[j_1,j_2-1]\cap S=[j_1+1,j_2]\cap T^L
 =\emptyset$, then $x_{[j_1,j_2]}$ is contained in a run of $\bm x$.
 \item[2)] If $[j'_1+1,j'_2]\cap S=
 [j'_1+1,j'_2]\cap T^L=\emptyset$, then $x'_{[j'_1,j'_2]}$ is
 contained in a run of $\bm x'$.
\end{itemize}
\end{lem}
\begin{proof}
We first prove 1). If $[j_1,j_2-1]\cap S=\emptyset$, then by the
definition of $S$, we have $x_{i}=x'_i$ for all $i\in[j_1,j_2-1]$;
if $[j_1+1,j_2]\cap T^L=\emptyset$, then by the definition of
$T^L$, we have $x_i=x'_{i-1}$ for all $i\in[j_1+1,j_2]$. Hence, we
can obtain $x_i=x_i'=x_{i+1}$ for all $i\in[j_1,j_2-1]$, which
implies that $x_{[j_1,j_2]}$ is contained in a run of $\bm x$.

The proof of 2) is similar to 1). From the assumption that
$[j'_1+1,j'_2]\cap S=[j'_1+1,j'_2]\cap T^L=\emptyset$, we can
obtain $x'_i=x_{i}=x'_{i-1}$ for all $i\in[j'_1+1,j'_2]$, which
implies that $x'_{[j'_1,j'_2]}$ is contained in a run of $\bm x'$.
\end{proof}

By Lemma \ref{iden-eq-dH}, if $j_1,j_2,j_1',j_2'$ satisfy the
conditions of Lemma \ref{iden-eq-dH}, then for any
$(j,j')\in[j_1,j_2]\times[j_1',j_2']$, we have $(x_{[n]\backslash
j},x'_{[n]\backslash j'})=(x_{[n]\backslash
j_\ell},x'_{[n]\backslash j_{\ell'}'})$ for any
$\ell,\ell'\in\{1,2\}$.


\subsection{The notation $\phi$}

For $a\in\Sigma_q$ and $j_1\neq j_2\in[n]$, let
$\phi^{j_1}_{j_2;a}(\bm x)$ be the sequence obtained from $\bm x$
by deleting $x_{j_1}$ and substituting $x_{j_2}$ with $a$. For
example, if $\bm x=10212201$, then $\phi^6_{3;0}(\bm x)=1001201$.

In our subsequent discussions, it will be helpful to describe
$B^{\text{S}}_{1}(x_{[n]\backslash j},x'_{[n]\backslash j'})$
using the notation $\phi$.

\begin{exam}
Suppose $\bm x=01010111$ and $\bm x'=01101011$. Then we have
$B^{\text{S}}_{1}(x_{[n]\backslash 4},x'_{[n]\backslash
7})=\{\phi^{4}_{7;x'_6}(\bm x), \phi^{4}_{3;x'_3}(\bm
x)\}=\{\phi^{7}_{3;x_3}(\bm x'), \phi^{7}_{6;x_7}(\bm x')\}$. In
fact, we can easily check that $x_{[n]\backslash 4}=0100111$ and
$\bm x'_{[n]\backslash 7}=0110101$. Moreover, we can find
$B^{\text{S}}_{1}(x_{[n]\backslash 4},x'_{[n]\backslash
7})=\{0100101, 0110111\}$, $\phi^{4}_{7;x'_6}(\bm
x)=0100101=\phi^{7}_{3;x_3}(\bm x')$ and $\phi^{4}_{3;x'_3}(\bm
x)=0110111=\phi^{7}_{6;x_7}(\bm x')$.
\end{exam}

In general, we have the following two remarks.
\begin{rem}\label{rem-phi-dH2}
For any $j,j'\in[n]$ such that $j\leq j'$, if
$d_{\text{H}}(x_{[n]\backslash j},x'_{[n]\backslash j'})=2$, then
by Observation 2, we can denote
$\{j_1,j_2\}=\big(S\cap[1,j-1]\big)\cup
\big(T^L\cap[j+1,j']\big)\cup\big(S\cap[j'+1,n]\big)$. For each
$\ell\in\{1,2\}$: if $j_\ell\in[1,j-1]\cup[j'+1,n]$, let $\bm
z_\ell=\phi^j_{j_\ell;x'_{j_\ell}}(\bm x)$ and $\bm
w_\ell=\phi^{j'}_{j_\ell;x_{j_\ell}}(\bm x')$; if
$j_\ell\in[j+1,j']$, let $\bm
z_\ell=\phi^j_{j_\ell;x'_{j_\ell-1}}(\bm x)$ and $\bm
w_\ell=\phi^{j'}_{j_{\ell}-1;x_{j_\ell}}(\bm x')$. Then we have
\begin{align*}
B^{\text{S}}_{1}(x_{[n]\backslash j},x'_{[n]\backslash j'})=\{\bm
z_1, \bm z_2\}=\{\bm w_1, \bm w_2\}.\end{align*} Similar results
can be obtained when $d_{\text{H}}(x_{[n]\backslash
j'},x'_{[n]\backslash j})=2$.
\end{rem}

\begin{rem}\label{rem-phi-dH1}
Similar to Remark \ref{rem-phi-dH2}, for any $j,j'\in[n]$ such
that $j\leq j'$ and $d_{\text{H}}(x_{[n]\backslash
j},x'_{[n]\backslash j'})=1$, then by Observation 2, we can denote
$\{j_1\}=\big(S\cap[1,j-1]\big)\cup
\big(T^L\cap[j+1,j']\big)\cup\big(S\cap[j'+1,n]\big)$. If
$j_1\in[1,j-1]\cup[j'+1,n]$, then we have \begin{align*}
B^{\text{S}}_{1}(x_{[n]\backslash j},x'_{[n]\backslash
j'})=\big\{\phi^j_{j_1;a}(\bm x):
a\in\Sigma_q\big\}=\big\{\phi^{j'}_{j_1;a}(\bm x'):
a\in\Sigma_q\big\};\end{align*} if $j_1\in[j+1,j']$, then we have
\begin{align*}
B^{\text{S}}_{1}(x_{[n]\backslash j},x'_{[n]\backslash
j'})=\big\{\phi^j_{j_1;a}(\bm x):
a\in\Sigma_q\big\}=\big\{\phi^{j'}_{j_1-1;a}(\bm x'):
a\in\Sigma_q\big\}.\end{align*} Similar results can be obtained
when $d_{\text{H}}(x_{[n]\backslash j'},x'_{[n]\backslash j})=1$.
\end{rem}

\section{Methodology}

In this section, we will always assume that $\bm x,\bm
x'\in\Sigma_q^n$ are arbitrarily chosen such that
$d=d_{\text{H}}(\bm x,\bm x')\geq 2$. By Observation 1, we have
$B^{\text{D,S}}_{1,1}(\bm x,\bm x')=\bigcup_{(\bm z,\bm
z')\in\Lambda}B^{\text{S}}_{1}(\bm z,\bm z')$, where
$\Lambda=\Lambda(\bm x,\bm x')$ is defined by \eqref{eq-def-LMD}.
To find the size of $B^{\text{D,S}}_{1,1}(\bm x,\bm x')$, we will
develop a method to divide the set $\Lambda$, and correspondingly
the set $B^{\text{D,S}}_{1,1}(\bm x,\bm x')$, into some subsets
that can be easily obtained from $\bm x$ and $\bm x'$. See Fig.
\ref{fig-divLMD} for an overview of the dividing of the set
$\Lambda$.

\begin{figure*}
\begin{center}
\includegraphics[width=13.9cm]{dct-code.1}
\end{center}
\vspace{-6pt}\caption{An overview of the dividing of $\Lambda$,
where $\Lambda=\big\{(x_{[n]\backslash j}, x'_{[n]\backslash j'}):
j,j'\in[n]~\text{and}~ d_{\text{H}}(x_{[n]\backslash j'},
x'_{[n]\backslash j})\leq 2\big\}$ is defined by
\eqref{eq-def-LMD}. First, $\Lambda$ is divided into $\Lambda_0,
\Lambda_1$ and $\Lambda_2$ according to the value of
$d_{\text{H}}(x_{[n]\backslash j'}, x'_{[n]\backslash j})$. Then
for each $\ell\in\{0,1,2\}$, $\Lambda_\ell$ is divided into
$\Lambda_\ell^L$ and $\Lambda_\ell^R$ according to the
relationship of $j$ and $j'$. Here we assume $j\leq j'$ and
consider $(x_{[n]\backslash j}, x'_{[n]\backslash j'})$ and
$(x_{[n]\backslash j'}, x'_{[n]\backslash j})$. Finally, for each
$\ell\in\{1,2\}$ and each $X\in\{L,R\}$, $\Lambda_\ell^X$ is
divided into $\Lambda_{\ell,i}^X$, $i=1,\cdots,p_\ell$, where
$p_1=3$ and $p_2=6$, according to the value of $(|S\cap
[1,j-1]|,|T^X\cap [j+1,j']|,|S\cap [j'+1,n]|)$, where by
Observation 2, $d_{\text{H}}(x_{[n]\backslash j'},
x'_{[n]\backslash j})=|S\cap [1,j-1]|+|T^X\cap [j+1,j']|+|S\cap
[j'+1,n]|$. Moreover, the sets $\Lambda^X_0$ and
$\Lambda^X_{\ell,i}$ can be easily obtained from $\bm x$ and $\bm
x'$.}\label{fig-divLMD}
\end{figure*}

\begin{defn}\label{defn-lmd-omg-ntn} For each
$\ell\in\{0,1,2\}$:
\begin{itemize}
 \item let
 \begin{align*}\Lambda_\ell=\Lambda_\ell(\bm x,\bm
 x')\triangleq\left\{(x_{[n]\backslash j}, x'_{[n]\backslash j'}):
 j,j'\in[n]~\text{and}~
 d_{\text{H}}(x_{[n]\backslash j'}, x'_{[n]\backslash
 j})=\ell\right\}\end{align*} and
 $$\Omega_{\ell}=\Omega_{\ell}(\bm x,\bm x')
 \triangleq\bigcup_{(\bm z,\bm
 z')\in\Lambda_{\ell}}
 B^{\text{S}}_{1}(\bm z,\bm z');$$
 \item let
 \begin{align*}\Lambda^L_\ell=\Lambda^L_\ell(\bm x,\bm
 x')&\triangleq\left\{(x_{[n]\backslash j}, x'_{[n]\backslash j'}):
 (j,j')\in[n]\times[n], j\leq j'~\text{and}~
 d_{\text{H}}(x_{[n]\backslash j}, x'_{[n]\backslash
 j'})=\ell\right\}\end{align*}
 and $$\Omega^L_{\ell}=\Omega^L_{\ell}(\bm x,\bm x')
 \triangleq\bigcup_{(\bm z,\bm
 z')\in\Lambda^L_{\ell}}
 B^{\text{S}}_{1}(\bm z,\bm z');$$
 \item let
 \begin{align*}\Lambda^R_\ell=\Lambda^R_\ell(\bm x,\bm
 x')&\triangleq\left\{(x_{[n]\backslash j'}, x'_{[n]\backslash j}):
 (j,j')\in[n]\times[n], j\leq j'~\text{and}~
 d_{\text{H}}(x_{[n]\backslash j'}, x'_{[n]\backslash
 j})=\ell\right\}\end{align*}
 and $$\Omega^R_{\ell}=\Omega^R_{\ell}(\bm x,\bm x')
 \triangleq\bigcup_{(\bm z,\bm
 z')\in\Lambda^R_{\ell}}
 B^{\text{S}}_{1}(\bm z,\bm z').$$
\end{itemize}
\end{defn}

By the above definitions, we have
$\Lambda_{\ell}=\Lambda^L_{\ell}\cup\Lambda^R_{\ell}$ for each
$\ell\in\{0,1,2\}$ and
$$\Lambda=\bigcup_{\ell=0}^2\Lambda_{\ell}
=\bigcup_{\ell=0}^2(\Lambda^L_{\ell}\cup\Lambda^R_{\ell}).$$
Correspondingly, we have
$\Omega_{\ell}=\Omega^L_{\ell}\cup\Omega^R_{\ell}$ for each
$\ell\in\{0,1,2\}$ and
$$B^{\text{D,S}}_{1,1}(\bm x,\bm
x')=\bigcup_{\ell=0}^2\Omega_{\ell}
=\bigcup_{\ell=0}^2(\Omega^L_{\ell}\cup\Omega^R_{\ell}).$$ Note
that $\Lambda^L_{\ell}$ and $\Lambda^R_{\ell}$ are not necessarily
disjoint, and so $\Omega^L_{\ell}$ and $\Omega^R_{\ell}$ are not
necessarily disjoint.

We remark that $(\bm x,\bm x')$ should be viewed as an ordered
pair in the notations $\Lambda^X_\ell(\bm x,\bm x')$,
$X\in\{L,R\}$ and $\ell\in\{0,1,2\}$. By the definitions, $(\bm
z,\bm z')\in\Lambda^R_\ell=\Lambda^R_\ell(\bm x,\bm x')$ if and
only if $(\bm z',\bm z)\in\Lambda^L_\ell(\bm x',\bm x)$.

In the following three subsections, we will determine the set
$\Lambda^X_\ell$ for each $X\in\{L,R\}$ and each
$\ell\in\{0,1,2\}$.

\subsection{For $\Lambda^L_0(\bm x,\bm x')$ and
$\Lambda^R_0(\bm x,\bm x')$}

We first consider $\Lambda^L_0(\bm x,\bm x')$. Let $S$ and $T^L$
be defined by \eqref{eq-def-S} and \eqref{eq-def-TL},
respectively. We have the following claim.

\begin{figure}[htbp]
\begin{center}
\includegraphics[width=8.0cm]{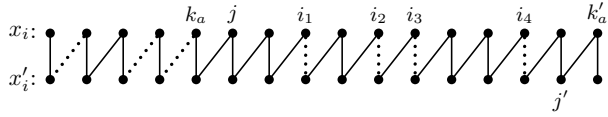}
\end{center}
\vspace{-6pt}\caption{An illustration of the pair $(j,j')$
satisfying \eqref{eq-LDM0-dst}. Each black dot represents a symbol
of $\bm x$ (in the upper row) or a symbol of $\bm x'$ (in the
lower row). Symbols are connected by a solid segment are
identical, while those connected by a dashed segment are distinct.
Here, $k_a=\max(T^L\cap[1,i_1])$ and $k'_a=n$ because
$T^L\cap[i_d+1,n]=\emptyset$. We can find that $(j,j')$ satisfies
\eqref{eq-LDM0-dst} if and only if $k_a\leq j\leq i_1<i_d\leq
j'\leq k'_a$. In this example, $d=4$. Moreover, we can find that
$x_i=x'_i=x_{i+1}$ for each $i\in [k_a,i_1-1]$ and
$x'_i=x_{i}=x'_{i-1}$ for each $i\in [i_d+1,k_a']$. Hence,
$x_{[k_a,i_1]}$ is contained in a run of $\bm x$ and
$x'_{[i_d,k'_a]}$ is contained in a run of $\bm
x'$.}\label{fig-LMD0}
\end{figure}

\textbf{Claim 0}: Suppose $d=d_{\text{H}}(\bm x,\bm x')\geq 2$.
\begin{itemize}
 \item[1)] If $|T^L\cap[i_1+1,i_d]|=0$, then $x_{[n]\backslash
 i_1}=x'_{[n]\backslash i_d}$ and
 $\Lambda^L_0=\{(x_{[n]\backslash
 i_1},x'_{[n]\backslash i_d})\}$. Hence, we have
 $\Omega^L_{0}=B^{\text{S}}_{1}(x_{[n]\backslash i_1},
 x'_{[n]\backslash i_d})
 =B^{\text{S}}_{1}(x_{[n]\backslash i_1})
 =B^{\text{S}}_{1}(x'_{[n]\backslash i_d})$.
 \item[2)] If $|T^L\cap[i_1+1,i_d]|\geq 1$, then
 $\Lambda^L_0(\bm x,\bm x')=\emptyset$.
\end{itemize}
\begin{proof}
Let $k_a=\max(T^L\cap[1,i_1])$ if $T^L\cap[1,i_1]\neq\emptyset$,
and $k_a=1$ otherwise. Similarly, let
$k'_a=\min(T^L\cap[i_d+1,n])-1$ if
$T^L\cap[i_d+1,n]\neq\emptyset$, and $k'_a=n$ otherwise. Then
$$k_a\leq i_1<i_d\leq k'_a.$$ By the definition, to
find $\Lambda^L_0(\bm x,\bm x')$, we need to find all
$(j,j')\in[n]\times[n]$ such that $j\leq j'$ and
$d_{\text{H}}(x_{[n]\backslash j},x'_{[n]\backslash j'})=0$. By
Observation 2, $d_{\text{H}}(x_{[n]\backslash j},x'_{[n]\backslash
j'})=0$ if and only if
$$|S\cap[1,j-1]|+|T^L\cap[j+1,j']|+|S\cap[j'+1,n]|=0,$$ or
equivalently,
\begin{align}\label{eq-LDM0-dst}
|S\cap[1,j-1]|=|T^L\cap[j+1,j']|=|S\cap[j'+1,n]|=0.\end{align}
Note that $j\leq j'$ and $S=\{i_1,\cdots,i_d\}$ such that
$i_1<\cdots<i_d$. Then from the conditions
$|S\cap[1,j-1]|=|S\cap[j'+1,n]|=0$, we have $j\leq i_1$ and
$j'\geq i_d$, which implies $$T^L\cap[i_1+1,i_d]\subseteq
T^L\cap[j+1,j'].$$ Combining this with the condition
$|T^L\cap[j+1,j']|=0$, we have
$$|T^L\cap[i_1+1,i_d]|\leq|T^L\cap[j+1,j']|=0.$$ Thus, if
$|T^L\cap[i_1+1,i_d]|\geq 1$, then there is no $(j,j')$ that
satisfies \eqref{eq-LDM0-dst}, and so $\Lambda^L_0(\bm x,\bm
x')=\emptyset$.

Conversely, if $|T^L\cap[i_1+1,i_d]|=0$, then clearly, we have
$x_{[n]\backslash i_1}=x'_{[n]\backslash i_d}$. Moreover, by the
definition of $k_a, k'_a, T^L$ and $S$, it is not hard to see that
$(j,j')$ satisfies \eqref{eq-LDM0-dst} if and only if $k_a\leq
j\leq i_1<i_d\leq j'\leq k'_a$ (see Fig. \ref{fig-LMD0}).
Therefore, we have $\Lambda^L_0(\bm x,\bm x')=\{(x_{[n]\backslash
j},x'_{[n]\backslash j'}): k_a\leq j\leq i_1<i_d\leq j'\leq
k'_a\}\neq\emptyset.$ Moreover, by the definition of $k_a$ and
$k'_a$, we can obtain $[k_a,i_1-1]\cap S=\emptyset$ and
$[k_a+1,i_1]\cap T^L=\emptyset$, so by 1) of Lemma
\ref{iden-eq-dH}, $x_{[k_a,i_1]}$ is contained in a run of $\bm
x$. Similarly, we can obtain $[i_d+1,k'_a]\cap S=[i_d+1,k'_a]\cap
T^L=\emptyset$, and so by 2) of Lemma \ref{iden-eq-dH},
$x'_{[i_d,k'_a]}$ is contained in a run of $\bm x'$. Thus, we have
$\Lambda^L_0(\bm x,\bm x')=\{(x_{[n]\backslash
j},x'_{[n]\backslash j'}): k_a\leq j\leq i_1<i_d\leq j'\leq
k'_a\}=\{(x_{[n]\backslash i_1},x'_{[n]\backslash i_d})\}$.
\end{proof}

For $\Lambda^R_0(\bm x,\bm x')$, let $T^R$ be defined according to
\eqref{eq-def-TR}, then we have the following claim.

\textbf{Claim 0$'$}: Suppose $d=d_{\text{H}}(\bm x,\bm x')\geq 2$.
\begin{itemize}
 \item[1)] If $|T^R\cap[i_1+1,i_d]|=0$, then $x_{[n]\backslash
 i_d}=x'_{[n]\backslash i_1}$ and
 $\Lambda^R_0=\{(x_{[n]\backslash
 i_d},x'_{[n]\backslash i_1})\}$. Hence, we have
 $\Omega^R_{0}=B^{\text{S}}_{1}(x_{[n]\backslash i_d},
 x'_{[n]\backslash i_1})
 =B^{\text{S}}_{1}(x_{[n]\backslash i_d})
 =B^{\text{S}}_{1}(x'_{[n]\backslash i_1})$.
 \item[2)] If $|T^R\cap[i_1+1,i_d]|\geq 1$, then
 $\Lambda^R_0(\bm x,\bm x')=\emptyset$.
\end{itemize}

Note that $T^R\cap[i_1+1,i_d]=T^R(\bm x,\bm
x')\cap[i_1+1,i_d]=T^L(\bm x',\bm x)\cap[i_1+1,i_d]$ and $(\bm
z,\bm z')\in\Lambda^R_0(\bm x,\bm x')$ if and only if $(\bm z',\bm
z)\in\Lambda^L_0(\bm x',\bm x)$. Also note that $\bm x,\bm
x'\in\Sigma_q^n$ are arbitrarily chosen. So, Claim $\bm{0'}$ can
be obtained directly from Claim $\bm{0}$.

\subsection{For $\Lambda^L_1(\bm x,\bm x')$ and
$\Lambda^R_1(\bm x,\bm x')$}

We first consider $\Lambda^L_1(\bm x,\bm x')$. By definition,
$\Lambda^L_1(\bm x,\bm x')$ is the set of all $(x_{[n]\backslash
j},x'_{[n]\backslash j'})$ such that $(j,j')\in[n]\times[n]$,
$j\leq j'$ and $d_{\text{H}}(x_{[n]\backslash j},x'_{[n]\backslash
j'})=1$. Then by Observation 2, we have
$|S\cap[1,j-1]|+|T^L\cap[j+1,j']|+|S\cap[j'+1,n]|=1$, and so there
are the following three cases to be considered.\vspace{2pt}

\noindent 1. $|S\cap[1,j-1]|=1$ and
$|T^L\cap[j+1,j']|=|S\cap[j'+1,n]|=0$.\vspace{2pt}

\noindent 2. $|T^L\cap[j+1,j']|=1$ and
$|S\cap[1,j-1]|=|S\cap[j'+1,n]|=0$.\vspace{2pt}

\noindent 3. $|S\cap[j'+1,n]|=1$ and
$|S\cap[1,j-1]|=|T^L\cap[j+1,j']|=0$.\vspace{4pt}

\noindent For each $i\in\{1,2,3\}$, let
$\Lambda^L_{1,i}=\Lambda^L_{1,i}(\bm x,\bm x')$ be the set of all
$(x_{[n]\backslash j},x'_{[n]\backslash j'})\in\Lambda^L_1(\bm
x,\bm x')$, where $(j,j')\in[n]\times[n]$ and $j\leq j'$, such
that the conditions of Case $i$ hold. Clearly, we have
$\Lambda^L_{1} =\bigcup_{i=1}^3\Lambda^L_{1,i}.$

If $T^L\cap[1,i_1]\neq\emptyset$, we let
\begin{align}\label{eq-def-k1}k_1=\max(T^L\cap[1,i_1]);\end{align}
if $T^L\cap[i_d+1,n]\neq\emptyset$, we let
\begin{align}\label{eq-def-k1p}
k'_1=\min(T^L\cap[i_d+1,n]).\end{align} Then
$$2\leq k_1\leq i_1<i_d<k'_1\leq n$$ and we have the
following Claims $1.1-1.3$.

\textbf{Claim 1.1}: If $T^L\cap [i_2+1,i_d]\neq\emptyset$, then
$\Lambda^L_{1,1}(\bm x,\bm x')=\emptyset$; if $T^L\cap
[i_2+1,i_d]=\emptyset$, then
$\Lambda^L_{1,1}=\big\{(x_{[n]\backslash i_2},x'_{[n]\backslash
i_d})\big\}$.

\textbf{Claim 1.2}: If $|T^L\cap [i_1+1,i_d]|\geq 2$, then
$\Lambda^L_{1,2}(\bm x,\bm x')=\emptyset$; if $|T^L\cap
[i_1+1,i_d]|=1$, then $\Lambda^L_{1,2}=\big\{(x_{[n]\backslash
i_1},x'_{[n]\backslash i_d})\big\}$; if $|T^L\cap [i_1+1,i_d]|=0$,
then we have $\Lambda^L_{1,2}\subseteq\big\{(x_{[n]\backslash
k_1-1},x'_{[n]\backslash i_d}), (x_{[n]\backslash
i_1},x'_{[n]\backslash k'_1})\big\}$.\footnote{Here
$\Lambda^L_{1,2}(\bm x,\bm x')\subseteq\big\{(x_{[n]\backslash
k_1-1},x'_{[n]\backslash i_d}), (x_{[n]\backslash
i_1},x'_{[n]\backslash k'_1})\big\}$ means that if $k_1~($resp.
$k'_1)$ exists, then $(x_{[n]\backslash k_1-1},x'_{[n]\backslash
i_d})\in\Lambda^L_{1,2}(\bm x,\bm x')~\big($resp.
$(x_{[n]\backslash i_1},x'_{[n]\backslash
k'_1})\in\Lambda^L_{1,2}(\bm x,\bm x')\big)$. The usage of the
notation $\subseteq$ in Claims 1.2$'$, 2.2, 2.4, 2.6, 2.2$'$,
2.4$'$ and 2.6$'$ should be understood similarly.}

\textbf{Claim 1.3}: If $T^L\cap [i_1+1,i_{d-1}]\neq\emptyset$,
then $\Lambda^L_{1,3}(\bm x,\bm x')=\emptyset$; if $T^L\cap
[i_1+1,i_{d-1}]=\emptyset$, then
$\Lambda^L_{1,3}=\big\{(x_{[n]\backslash i_1},x'_{[n]\backslash
i_{d-1}})\big\}$.

Similarly, we can divide $\Lambda^R_{1}=\Lambda^R_{1}(\bm x,\bm
x')$ into three subsets $\Lambda^R_{1,i}=\Lambda^R_{1,i}(\bm x,\bm
x')$, $i=1,2,3$, according to the value of $(|S\cap
[1,j-1]|,|T^R\cap [j+1,j']|,|S\cap [j'+1,n]|)~($see Table 1), and
we can obtain $\Lambda^R_{1} =\bigcup_{i=1}^3\Lambda^R_{1,i}.$ If
$T^R\cap[1,i_1]\neq\emptyset$, let
\begin{align}\label{eq-def-bk1}
m_1=\max(T^R\cap[1,i_1]);\end{align} if
$T^R\cap[i_d+1,n]\neq\emptyset$, let
\begin{align}\label{eq-def-bk1p}
m'_1=\min(T^R\cap[i_d+1,n]).\end{align} Then
$$2\leq m_1\leq i_1<i_d< m'_1\leq n$$
and we have the following Claims 1.1$'-$1.3$'$.

\textbf{Claim 1.1$\bm '$}: If $T^R\cap [i_2+1,i_d]\neq\emptyset$,
then $\Lambda^R_{1,1}(\bm x,\bm x')=\emptyset$; if $T^R\cap
[i_2+1,i_d]=\emptyset$, then
$\Lambda^R_{1,1}=\big\{(x_{[n]\backslash i_d},x'_{[n]\backslash
i_2})\big\}$.

\textbf{Claim 1.2$\bm '$}: If $|T^R\cap [i_1+1,i_d]|\geq 2$, then
$\Lambda^R_{1,2}(\bm x,\bm x')=\emptyset$; if $|T^R\cap
[i_1+1,i_d]|=1$, then $\Lambda^R_{1,2}=\big\{(x_{[n]\backslash
i_d},x'_{[n]\backslash i_1})\big\}$; if $|T^R\cap [i_1+1,i_d]|=0$,
then we have $\Lambda^R_{1,2}(\bm x,\bm
x')\subseteq\big\{(x_{[n]\backslash i_d,x'_{[n]\backslash
m_1-1}}), (x_{[n]\backslash  m'_1},x'_{[n]\backslash i_1})\big\}$.

\textbf{Claim 1.3$\bm '$}: If $T^R\cap
[i_1+1,i_{d-1}]\neq\emptyset$, then $\Lambda^R_{1,3}(\bm x,\bm
x')=\emptyset$; if $T^R\cap [i_1+1,i_{d-1}]=\emptyset$, then
$\Lambda^R_{1,3}=\big\{(x_{[n]\backslash
i_{d-1}},x'_{[n]\backslash i_1})\big\}$.

\begin{table}
\begin{center}
\renewcommand\arraystretch{1.5}
\begin{tabular}{|c|c|c|c|c|c|}
\hline $~$ & $|S\cap [1,j-1]|$ & $|T^X\cap [j+1,j']|$ & $|S\cap
[j'+1,n]|$ \\
\hline $\Lambda^X_{0}$   & $0$ & $0$  & $0$ \\
\hline $\Lambda^X_{1,1}$ & $1$ & $0$  & $0$ \\
\hline $\Lambda^X_{1,2}$ & $0$ & $1$  & $0$ \\
\hline $\Lambda^X_{1,3}$ & $0$ & $0$  & $1$ \\
\hline $\Lambda^X_{2,1}$ & $2$ & $0$  & $0$ \\
\hline $\Lambda^X_{2,2}$ & $0$ & $2$  & $0$ \\
\hline $\Lambda^X_{2,3}$ & $0$ & $0$  & $2$ \\
\hline $\Lambda^X_{2,4}$ & $0$ & $1$  & $1$ \\
\hline $\Lambda^X_{2,5}$ & $1$ & $0$  & $1$ \\
\hline $\Lambda^X_{2,6}$ & $1$ & $1$  & $0$ \\
\hline
\end{tabular}\\
\end{center}
\vspace{2.5mm} Table 1. For each $X\in\{L,R\}$ and each
$\ell\in\{0,1,2\}$, by Definition \ref{defn-lmd-omg-ntn} and
Observation 2, the set $\Lambda^X_\ell$ can be determined by the
tuple $(|S\cap [1,j-1]|,|T^X\cap [j+1,j']|,|S\cap [j'+1,n]|)$ for
each $(j,j')\in[n]\times[n]$ such that $j\leq j'$. Moreover, the
set $\Lambda^X_1$ is divided into three subsets $\Lambda^X_{1,i}$,
$i=1,2,3$, and the set $\Lambda^X_2$ is divided into six subsets
$\Lambda^X_{2,i}$, $i=1,2,\cdots,6$, according to the value of
$(|S\cap [1,j-1]|,|T^X\cap [j+1,j']|,|S\cap [j'+1,n]|)$.
\end{table}

\begin{rem}\label{rem-LMD1-to-OMG1}
For each $X\in\{L,R\}$ and $i\in\{1,2,3\}$, let
$$\Omega^X_{1,i}=\Omega^X_{1,i}(\bm x,\bm x')
\triangleq\bigcup_{(\bm z,\bm z')\in\Lambda^X_{1,i}}
B^{\text{S}}_{1}(\bm z,\bm z').$$ Then we have
$\Omega^X_{1}=\bigcup_{i=1}^3\Omega^X_{1,i}.$ Moreover, we can
easily obtain $\Omega^X_{1,i}$ from $\Lambda^X_{1,i}$ by Remark
\ref{rem-phi-dH1}. As an example, consider $\Omega^R_{1,2}$ with
$|T^R\cap [i_1+1,i_d]|=1$. By Claim 1.2$'$, we have
$\Lambda^R_{1,2}=\{(x_{[n]\backslash i_d},x'_{[n]\backslash
i_1})\}$. Let $\{j_1'\}=T^R\cap [i_1+1,i_d]$. Then by Remark
\ref{rem-phi-dH1}, we can obtain
$\Omega^R_{1,2}=B^{\text{S}}_{1}(x_{[n]\backslash
i_d},x'_{[n]\backslash i_1})=\{\phi^{i_d}_{j_1'-1;a}(\bm x):
a\in\Sigma_q\}=\{\phi^{i_1}_{j_1';a}(\bm x'): a\in\Sigma_q\}$. In
particular, we have $|B^{\text{S}}_{1}(\bm z,\bm z')|=q$ for each
$(\bm z,\bm z')\in\Lambda^X_{1}$ and each $X\in\{L,R\}$.
\end{rem}

In the following, we prove Claims 1.1$-$1.3. Note that
$T^R=T^R(\bm x,\bm x')=T^L(\bm x',\bm x)$ and $(\bm z,\bm
z')\in\Lambda^R_1(\bm x,\bm x')$ if and only if $(\bm z',\bm
z)\in\Lambda^L_1(\bm x',\bm x)$. So, Claims 1.1$'-$1.3$'$ can be
obtained directly from Claims 1.1$-$1.3.

Let
\begin{equation}\label{defeq-ka-1}
k_b=\!\left\{\!\begin{aligned} &\max(T^L\cap[1,i_1]\backslash
k_1),
~\text{if}~|T^L\cap[1,i_1]|\geq 2;\\
&1, ~\text{otherwise}.
\end{aligned}\right.
\end{equation}
and
\begin{equation}\label{defeq-pka-1}
k'_b=\!\left\{\!\begin{aligned}
&\min(T^L\cap[i_d+1,n]\backslash\{k'_1\})-1,
~\text{if}~|T^L\cap[i_d+1,n]|\geq 2;\\
&n, ~\text{otherwise}.
\end{aligned}\right.
\end{equation}
Then we have
\begin{align}\label{eq-LMD1-all-order}
k_b<k_1\leq i_1<i_d<k'_1\leq k'_b.\end{align}

\begin{figure}[htbp]
\begin{center}
\includegraphics[width=8.0cm]{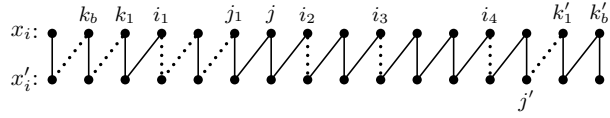}
\end{center}
\vspace{-6pt}\caption{An illustration of the pair $(j,j')$ in the
proof of Claim 1.1. Here, $S=\{i_1,i_2,i_3,i_4\}$, $k_1$ and
$k_1'$ are defined by \eqref{eq-def-k1} and \eqref{eq-def-k1p}
respectively. According to the proof of Claim 1.1,
$j_1=\max(T^L\cap[i_1+1,i_2])$. We can see that $(j,j')$ satisfies
the conditions $|S\cap[1,j-1]|=1$ and
$|T^L\cap[j+1,j']|=|S\cap[j'+1,n]|=0$ if and only if it satisfies
\eqref{eq-dHGE2-LMD1-L2-Clm1}, that is, $j_1\leq j\leq i_2\leq
i_d\leq j'<k'_1$. In fact, we have
$S\cap[1,j-1]=S\cap[1,j-1]=\{i_1\}$ and
$T^L\cap[j+1,j']=S\cap[j'+1,n]=\emptyset$. Moreover, we can see
that $x_{[j_1,i_2]}$ is contained in a run of $\bm x$ and
$x'_{[i_d,k'_1-1]}$ is contained in a run of $\bm
x'$.}\label{fig-LMD1-1}
\end{figure}

\begin{proof}[Proof of Claim 1.1]
By definition, $\Lambda^L_{1,1}(\bm x,\bm x')\neq\emptyset$ if and
only if there exist $(j,j')$ satisfying conditions of Case 1
$($i.e., $|S\cap[1,j-1]|=1$ and
$|T^L\cap[j+1,j']|=|S\cap[j'+1,n]|=0)$.

Suppose $(j,j')$ satisfying the conditions of Case 1. Then from
$|S\cap[1,j-1]|=1$ and $|S\cap[j'+1,n]|=0$, we have $i_1<j\leq
i_2\leq i_d\leq j'.$ Combining this with $|T^L\cap[j+1,j']|=0$
(condition of Case 1), we obtain $T^L\cap [i_2+1,i_d]\subseteq
T^L\cap [j+1,j']=\emptyset$. Hence, if $T^L\cap
[i_2+1,i_d]\neq\emptyset$, then we have $\Lambda^L_{1,1}(\bm x,\bm
x')=\emptyset$.

Conversely, suppose $T^L\cap [i_2+1,i_d]=\emptyset$. We need to
prove that $\Lambda^L_{1,1}(\bm x,\bm x')=\{(x_{[n]\backslash
i_2},x'_{[n]\backslash i_d})\}$. Let
$j_1=\max(T^L\cap[i_1+1,i_2])$ if
$T^L\cap[i_1+1,i_2]\neq\emptyset$, and $j_1=i_1+1$ otherwise. Then
by \eqref{eq-LMD1-all-order}, we have $i_1<j_1\leq i_2\leq
i_d<k'_1.$ It is not hard to verify that $(j,j')$ satisfies the
conditions of Case 1 $($i.e., $|S\cap[1,j-1]|=1$ and
$|T^L\cap[j+1,j']|=|S\cap[j'+1,n]|=0)$ if and only if (see Fig.
\ref{fig-LMD1-1})
\begin{align}\label{eq-dHGE2-LMD1-L2-Clm1}
j_1\leq j\leq i_2\leq i_d\leq j'<k'_1.\end{align} Therefore,
$\Lambda^L_{1,1}(\bm x,\bm x')=\{(x_{[n]\backslash
j},x'_{[n]\backslash j'}): j_1\leq j\leq i_2\leq i_d\leq
j'<k'_1\}\neq\emptyset$. Note that by
\eqref{eq-dHGE2-LMD1-L2-Clm1} and by the definition of $j_1$ and
$k'_1$, we can obtain $[j_1+1,i_2]\cap T^L=\emptyset$ and
$[i_d+1,k'_1-1]\cap T^L=\emptyset.$ Moreover, by the definition of
$S$ and $j_1$, we can obtain $[j_1,i_2-1]\cap S=\emptyset$ and
$[i_d+1,k'_1-1]\cap S=\emptyset.$ Hence, by Lemma
\ref{iden-eq-dH}, $x_{[j_1,i_2]}~($resp. $x'_{[i_d,k'_1-1]})$ is
contained in a run of $\bm x~($resp. $\bm x')$, which implies that
$\Lambda^L_{1,1}(\bm x,\bm x')=\{(x_{[n]\backslash
j},x'_{[n]\backslash j'}): j_1\leq j\leq i_2\leq i_d\leq
j'<k'_1\}=\{(x_{[n]\backslash i_2},x'_{[n]\backslash i_d})\}$.
\end{proof}

\begin{figure}[htbp]
\begin{center}
\includegraphics[width=8.0cm]{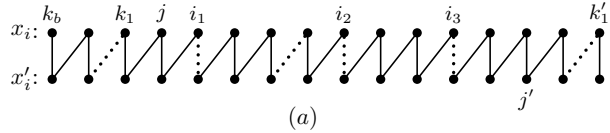}\vspace{12pt}\\
\includegraphics[width=8.0cm]{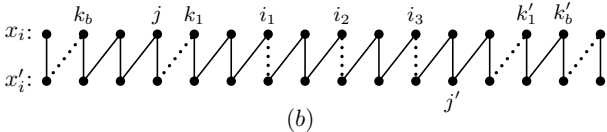}~~~~~~~~~~~~
\includegraphics[width=8.0cm]{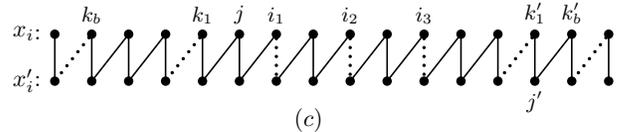}
\end{center}
\vspace{-6pt}\caption{An illustration of the pair $(j,j')$ in the
proof of Claim 1.2. Here $S=\{i_1,i_2,i_3\}$. In this figure, (a)
is for Case (i), (b) is for possibility 1) of Case (ii) and (c) is
for possibility 2) of Case (ii). Here, $k_1,k_1',k_b$ and $k_b'$
are defined by \eqref{eq-def-k1}, \eqref{eq-def-k1p},
\eqref{defeq-ka-1} and \eqref{defeq-pka-1}
respectively.}\label{fig-LMD1-2}
\end{figure}

\begin{proof}[Proof of Claim 1.2]
By definition, $\Lambda^L_{1,2}(\bm x,\bm x')\neq\emptyset$ if and
only if there exist $(j,j')$ satisfying conditions of Case 2
$($i.e., $|S\cap[1,j-1]|=1$ and
$|T^L\cap[j+1,j']|=|S\cap[j'+1,n]|=0)$.

Suppose $(j,j')$ satisfying the conditions of Case 2. By the
condition $|S\cap[1,j-1]|=|S\cap[j'+1,n]|=0$, we have $j\leq
i_1<i_d\leq j'.$ Combining this with $|T^L\cap[j+1,j']|=1$
(condition of Case 2), we have $|T^L\cap [i_1+1,i_d]|\leq
|T^L\cap[j+1,j']|=1$. Hence, $\Lambda^L_{1,2}(\bm x,\bm
x')=\emptyset$ if $|T^L\cap [i_1+1,i_d]|\geq 2$.

Conversely, suppose $|T^L\cap [i_1+1,i_d]|\leq 1$. We need to
consider the following Cases (i) and (ii).

Case (i): $|T^L\cap [i_1+1,i_d]|=1$. Then by the conditions of
Case 2, and by \eqref{eq-LMD1-all-order}, we have $k_1\leq j\leq
i_1<i_d\leq j'<k'_1$ (see Fig. \ref{fig-LMD1-2} (a)). Similar to
Claim 1.1, we can prove $x_{[k_1,i_1]}~($resp.
$x'_{[i_d,k'_1-1]})$ is contained in a run of $\bm x~($resp. $\bm
x')$, so we have $\Lambda^L_{1,2}(\bm x,\bm
x')=\{(x_{[n]\backslash j},x'_{[n]\backslash j'}): k_1\leq j\leq
i_1<i_d\leq j'<k'_1\}=\{(x_{[n]\backslash i_1},x'_{[n]\backslash
i_d})\}$.

Case (ii): $|T^L\cap [i_1+1,i_d]|=0$. In this case, by conditions
of Case 2, and by \eqref{eq-LMD1-all-order}, we have two
possibilities: 1) $k_b\leq j<k_1\leq i_1<i_d\leq j'<k'_1$ (see
Fig. \ref{fig-LMD1-2} (b)), which implies $(x_{[n]\backslash
k_1-1},x'_{[n]\backslash i_d})\in\Lambda^L_{1,2}(\bm x,\bm x')$;
and 2) $k_1\leq j\leq i_1<i_d<k'_1\leq j'\leq k'_b$ (see Fig.
\ref{fig-LMD1-2} (c)), which implies $(x_{[n]\backslash
i_1},x'_{[n]\backslash k'_1})\in\Lambda^L_{1,2}(\bm x,\bm x')$.
Thus, we have $\Lambda^L_{1,2}(\bm x,\bm
x')\subseteq\{(x_{[n]\backslash k_1-1},x'_{[n]\backslash i_d}),
(x_{[n]\backslash i_1},x'_{[n]\backslash k'_1})\}$.
\end{proof}

\begin{figure}[htbp]
\begin{center}
\includegraphics[width=8.0cm]{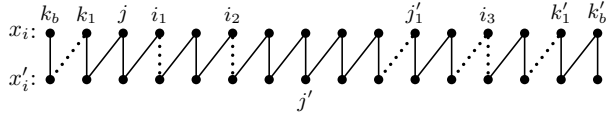}
\end{center}
\vspace{-6pt}\caption{An illustration of the pair $(j,j')$ in the
proof of Claim 1.3. Here, $S=\{i_1,i_2,i_3\}$ and $j_1'
=\min(T^L\cap[i_{2}+1,i_3])$ because $d=3$.}\label{fig-LMD1-3}
\end{figure}

\begin{proof}[Proof of Claim 1.3]
The proof is similar to Claim 1.1.

First suppose $(j,j')$ satisfies conditions of Case 3 $($i.e.,
$|S\cap[j'+1,n]|=1$ and $|S\cap[1,j-1]|=|T^L\cap[j+1,j']|=0)$.
Then by $|S\cap[1,j-1]|=0$ and $|S\cap[j'+1,n]|=1$, we have
$$j\leq i_1\leq i_{d-1}\leq j'<i_d.$$ Combining this with
$|T^L\cap[j+1,j']|=0$, we have $T^L\cap [i_1+1,i_{d-1}]\subseteq
T^L\cap [j+1,j']=\emptyset$, which implies that
$\Lambda^L_{1,3}(\bm x,\bm x')=\emptyset$ if $T^L\cap
[i_1+1,i_{d-1}]\neq\emptyset$.

Conversely, suppose $T^L\cap [i_1+1,i_{d-1}]=\emptyset$. Let
$j_1'=\min(T^L\cap[i_{d-1}+1,i_d])$ if
$T^L\cap[i_{d-1}+1,i_d]\neq\emptyset$, and $j_1'=i_d$ otherwise.
Then by \eqref{eq-LMD1-all-order}, we have
$$k_1\leq i_1\leq i_{d-1}<j'_1\leq i_d.$$ Clearly, $(j,j')$ satisfies
the conditions of Case 3 $($i.e., $|S\cap[j'+1,n]|=1$ and
$|S\cap[1,j-1]|=|T^L\cap[j+1,j']|=0)$ if and only if (see Fig.
\ref{fig-LMD1-3})
\begin{align}\label{eq-dHGE2-LMD1-L2-Clm3}
k_1\leq j\leq i_1\leq i_{d-1}\leq j'<j'_1.\end{align} By Lemma
\ref{iden-eq-dH}, we can prove $x_{[k_1,i_1]}~($resp.
$x'_{[i_{d-1},j'_1-1]})$ is contained in a run of $\bm x~($resp.
$\bm x')$, which implies that $\Lambda^L_{1,3}(\bm x,\bm
x')=\{(x_{[n]\backslash j},x'_{[n]\backslash j'}): k_1\leq j\leq
i_1\leq i_{d-1}\leq j'<j'_1\}=\{(x_{[n]\backslash
i_1},x'_{[n]\backslash i_{d-1}})\}$.
\end{proof}

\subsection{For $\Lambda^L_2(\bm x,\bm x')$ and
$\Lambda^R_2(\bm x,\bm x')$}

We first consider $\Lambda^L_2(\bm x,\bm x')$. Recall that
$\Lambda^L_2(\bm x,\bm x')$ is the set of all $(x_{[n]\backslash
j},x'_{[n]\backslash j'})$ such that $(j,j')\in[n]\times[n]$,
$j\leq j'$ and $d_{\text{H}}(x_{[n]\backslash j},x'_{[n]\backslash
j'})=2$. By Observation 2, we have
$|S\cap[1,j-1]|+|T^L\cap[j+1,j']|+|S\cap[j'+1,n]|=2$, and we need
to consider the following six cases.\vspace{2pt}

\noindent 1. $|S\cap[1,j-1]|=2$ and
$|T^L\cap[j+1,j']|=|S\cap[j'+1,n]|=0$.\vspace{2pt}

\noindent 2. $|T^L\cap[j+1,j']|=2$ and
$|S\cap[1,j-1]|=|S\cap[j'+1,n]|=0$.\vspace{2pt}

\noindent 3. $|S\cap[j'+1,n]|=2$ and
$|S\cap[1,j-1]|=|T^L\cap[j+1,j']|=0$.\vspace{2pt}

\noindent 4. $|S\cap[1,j-1]|=0$ and
$|T^L\cap[j+1,j']|=|S\cap[j'+1,n]|=1$.\vspace{2pt}

\noindent 5. $|T^L\cap[j+1,j']|=0$ and
$|S\cap[1,j-1]|=|S\cap[j'+1,n]|=1$.\vspace{2pt}

\noindent 6. $|S\cap[j'+1,n]|=0$ and
$|S\cap[1,j-1]|=|T^L\cap[j+1,j']|=1$.\vspace{4pt}

\noindent For each $i\in\{1,2,\cdots,6\}$, let
$\Lambda^L_{2,i}=\Lambda^L_{2,i}(\bm x,\bm x')$ be the set of
$(x_{[n]\backslash j}, x'_{[n]\backslash j'})\in\Lambda^L_2(\bm
x,\bm x')$ such that $(j,j')\in[n]\times[n]$, $j\leq j'$ and
conditions of Case $i$ hold. Then
$\Lambda^L_{2}=\bigcup_{i=1}^6\Lambda^L_{2,i}.$ If
$|T^L\cap[1,i_1]|\geq 2$, let
\begin{align}\label{eq-def-k2}
k_2=\max(T^L\cap[1,i_1]\backslash k_1)\end{align} where $k_1$ is
defined as in \eqref{eq-def-k1}; if $|T^L\cap[i_d+1,n]|\geq 2$,
let
\begin{align}\label{eq-def-k2p}
k'_2=\min(T^L\cap[i_d+1,n]\backslash k'_1)\end{align} where $k'_1$
is defined as in \eqref{eq-def-k1p}. Then
$$2\leq k_2<k_1\leq i_1<i_d<k'_1<k'_2\leq
n$$ and we have the following Claims 2.1$-$2.6.

\textbf{Claim 2.1}: If $d=2$, then
$\Lambda^L_{2,1}=\{(x_{[n]\backslash j},x'_{[n]\backslash j}):
j\in[i_2+1,n]\}$ and $|\Lambda^L_{2,1}|$ equals to the number of
runs of $x_{[i_2+1,n]}$; if $d\geq 3$ and $T^L\cap
[i_3+1,i_d]\neq\emptyset$, then $\Lambda^L_{2,1}=\emptyset$; if
$d\geq 3$ and $T^L\cap [i_3+1,i_d]=\emptyset$, then
$\Lambda^L_{2,1}=\{(x_{[n]\backslash i_3},x'_{[n]\backslash
i_d})\}$.

\textbf{Claim 2.2}: If $|T^L\cap [i_1+1,i_d]|\geq 3$, then we have
$\Lambda^L_{2,2}(\bm x,\bm x')=\emptyset$; if $|T^L\cap
[i_1+1,i_d]|=2$, then $\Lambda^L_{2,2}=\{(x_{[n]\backslash
i_1},x'_{[n]\backslash i_d})\}$; if $|T^L\cap [i_1+1,i_d]|=1$,
then $\Lambda^L_{2,2}\subseteq\{(x_{[n]\backslash
k_1-1},x'_{[n]\backslash i_d}), (x_{[n]\backslash
i_1},x'_{[n]\backslash k'_1})\}$; if $|T^L\cap [i_1+1,i_d]|=0$,
then $\Lambda^L_{2,2}\subseteq\{(x_{[n]\backslash
k_2-1},x'_{[n]\backslash i_d}), (x_{[n]\backslash
k_1-1},x'_{[n]\backslash k'_1}), (x_{[n]\backslash
i_1},x'_{[n]\backslash k'_2})\}$.

\textbf{Claim 2.3}: If $d=2$, then
$\Lambda^L_{2,3}=\{(x_{[n]\backslash j},x'_{[n]\backslash j}):
j\in[1,i_1-1]\}$ and $|\Lambda^L_{2,3}|$ equals to the number of
runs of $x_{[1,i_1-1]}$; if $d\geq 3$ and $T^L\cap
[i_1+1,i_{d-2}]\neq\emptyset$, then $\Lambda^L_{2,3}=\emptyset$;
if $d\geq 3$ and $T^L\cap [i_1+1,i_{d-2}]=\emptyset$, then
$\Lambda^L_{2,3}=\{(x_{[n]\backslash i_1},x'_{[n]\backslash
i_{d-2}})\}$.

\textbf{Claim 2.4}: If $|T^L\cap [i_1+1,i_{d-1}]|\geq 2$, then
$\Lambda^L_{2,4}=\emptyset$; if $|T^L\cap [i_1+1,i_{d-1}]|=1$,
then $\Lambda^L_{2,4}=\{(x_{[n]\backslash i_1},x'_{[n]\backslash
i_{d-1}})\}$; if $|T^L\cap[i_1+1,i_{d-1}]|=0$, then
$\Lambda^L_{2,4}\subseteq\{(x_{[n]\backslash
k_1-1},x'_{[n]\backslash i_{d-1}}), (x_{[n]\backslash
i_1},x'_{[n]\backslash j_1'})\}$, where
$j_1'=\min(T^L\cap[i_{d-1}+1,i_{d}-1])$ when
$T^L\cap[i_{d-1}+1,i_{d}-1]\neq\emptyset$.

\textbf{Claim 2.5}: If $d=2$, then
$\Lambda^L_{2,5}=\{(x_{[n]\backslash j},x'_{[n]\backslash j}):
j\in[i_1+1,i_2-1]\}$ and $|\Lambda^L_{2,5}|$ equals to the number
of runs of $x_{[i_1+1,i_2-1]}$; if $d\geq 3$ and
$T^L\cap[i_2+1,i_{d-1}]\neq\emptyset$, then
$\Lambda^L_{2,5}=\emptyset$; if $d\geq 3$ and
$T^L\cap[i_2+1,i_{d-1}]=\emptyset$, then
$\Lambda^L_{2,5}=\{(x_{[n]\backslash i_2},x'_{[n]\backslash
i_{d-1}})\}$.

\textbf{Claim 2.6}: If $|T^L\cap [i_2+1,i_{d}]|\geq 2$, then
$\Lambda^L_{2,6}=\emptyset$; if $|T^L\cap [i_2+1,i_{d}]|=1$, then
$\Lambda^L_{2,6}=\{(x_{[n]\backslash i_2},x'_{[n]\backslash
i_{d}})\}$; if $|T^L\cap[i_2+1,i_{d}]|=0$, then
$\Lambda^L_{2,6}\subseteq\{(x_{[n]\backslash
i_2},x'_{[n]\backslash k'_{1}}), (x_{[n]\backslash
j_1-1},x'_{[n]\backslash i_{d}})\}$, where
$j_1=\max(T^L\cap[i_{1}+2,i_{2}])$ when
$T^L\cap[i_{1}+2,i_{2}]\neq\emptyset$.

Similarly, the set $\Lambda^R_{2}=\Lambda^R_{2}(\bm x,\bm x')$ can
be divided into six subsets $\Lambda^R_{2,i}=\Lambda^R_{2,i}(\bm
x,\bm x')$, $i=1,2,\cdots,6$, according to the value of $(|S\cap
[1,j-1]|,|T^R\cap [j+1,j']|,|S\cap [j'+1,n]|)~($see table 1$)$,
and $\Lambda^R_{2} =\bigcup_{i=1}^6\Lambda^R_{2,i}.$ If
$|T^L\cap[1,i_1]|\geq 2$, let
\begin{align}\label{eq-def-bk2}
 m_2=\max(T^R\cap[1,i_1]\backslash  m_1)\end{align}
where $ m_1$ is defined as in \eqref{eq-def-bk1}; if
$|T^R\cap[i_d+1,n]|\geq 2$, let
\begin{align}\label{eq-def-bk2p}
 m'_2=\min(T^R\cap[i_d+1,n]\backslash  m'_1)\end{align}
where $ m'_1$ is defined as in \eqref{eq-def-bk1p}. Then
$$2\leq m_2< m_1\leq i_1<i_d< m'_1< m'_2\leq n$$
and we have the following Claims 2.1$'-$2.6$'$.

\textbf{Claim 2.1$\bm '$}: If $d=2$, then
$\Lambda^R_{2,1}=\{(x_{[n]\backslash j},x'_{[n]\backslash j}):
j\in[i_2+1,n]\}$ and $|\Lambda^R_{2,1}|$ equals to the number of
runs of $x_{[i_2+1,n]}$; if $d\geq 3$ and $T^R\cap
[i_3+1,i_d]\neq\emptyset$, then $\Lambda^R_{2,1}=\emptyset$; if
$d\geq 3$ and $T^R\cap [i_3+1,i_d]=\emptyset$, then
$\Lambda^R_{2,1}=\{(x_{[n]\backslash i_d},x'_{[n]\backslash
i_3})\}$.

\textbf{Claim 2.2$\bm '$}: If $|T^R\cap [i_1+1,i_d]|\geq 3$, then
we have $\Lambda^R_{2,2}=\emptyset$; if $|T^R\cap [i_1+1,i_d]|=2$,
then $\Lambda^R_{2,2}=\{(x_{[n]\backslash i_d},x'_{[n]\backslash
i_1})\}$; if $|T^R\cap [i_1+1,i_d]|=1$, then
$\Lambda^R_{2,2}\subseteq\{(x_{[n]\backslash
i_d},x'_{[n]\backslash
 m_1-1}), (x_{[n]\backslash  m'_1},x'_{[n]\backslash
i_1})\}$; if $|T^R\cap [i_1+1,i_d]|=0$, then
$\Lambda^R_{2,2}\subseteq\{(x_{[n]\backslash
i_d},x'_{[n]\backslash
 m_2-1}), (x_{[n]\backslash  m'_1},x'_{[n]\backslash
 m_1-1}), (x_{[n]\backslash  m'_2},x'_{[n]\backslash
i_1})\}$.

\textbf{Claim 2.3$\bm '$}: If $d=2$, then
$\Lambda^R_{2,3}=\{(x_{[n]\backslash j},x'_{[n]\backslash j}):
j\in[1,i_1-1]\}$ and $|\Lambda^R_{2,3}|$ equals to the number of
runs of $x_{[1,i_1-1]}$; if $d\geq 3$ and $T^R\cap
[i_1+1,i_{d-2}]\neq\emptyset$, then $\Lambda^R_{2,3}=\emptyset$;
if $d\geq 3$ and $T^R\cap [i_1+1,i_{d-2}]=\emptyset$, then
$\Lambda^R_{2,3}=\{(x_{[n]\backslash i_{d-2}},x'_{[n]\backslash
i_{1}})\}$.

\textbf{Claim 2.4$\bm '$}: If $|T^R\cap [i_1+1,i_{d-1}]|\geq 2$,
then $\Lambda^R_{2,4}=\emptyset$; if $|T^R\cap
[i_1+1,i_{d-1}]|=1$, then $\Lambda^R_{2,4}=\{(x_{[n]\backslash
i_{d-1}},x'_{[n]\backslash i_{1}})\}$; if
$|T^R\cap[i_1+1,i_{d-1}]|=0$, then
$\Lambda^R_{2,4}\subseteq\{(x_{[n]\backslash
i_{d-1}},x'_{[n]\backslash m_1-1}), (x_{[n]\backslash
\bar{j}_1'},x'_{[n]\backslash i_1})\}$, where
$\bar{j}_1'=\min(T^R\cap[i_{d-1}+1,i_{d}-1])$ when
$T^R\cap[i_{d-1}+1,i_{d}-1]\neq\emptyset$.

\textbf{Claim 2.5$\bm '$}: If $d=2$, then
$\Lambda^R_{2,5}=\{(x_{[n]\backslash j},x'_{[n]\backslash j}):
j\in[i_1+1,i_2-1]\}$ and $|\Lambda^R_{2,5}|$ equals to the number
of runs of $x_{[i_1+1,i_2-1]}$; if $d\geq 3$ and
$T^R\cap[i_2+1,i_{d-1}]\neq\emptyset$, then
$\Lambda^R_{2,5}=\emptyset$; if $d\geq 3$ and
$T^R\cap[i_2+1,i_{d-1}]=\emptyset$, then
$\Lambda^R_{2,5}=\{(x_{[n]\backslash i_{d-1}},x'_{[n]\backslash
i_{2}})\}$.

\textbf{Claim 2.6$\bm '$}: If $|T^R\cap [i_2+1,i_{d}]|\geq 2$,
then $\Lambda^R_{2,6}=\emptyset$; if $|T^R\cap [i_2+1,i_{d}]|=1$,
then $\Lambda^R_{2,6}=\{(x_{[n]\backslash i_d},x'_{[n]\backslash
i_{2}})\}$; if $|T^R\cap[i_2+1,i_{d}]|=0$, then
$\Lambda^R_{2,6}\subseteq\{(x_{[n]\backslash m'_{1}},
x'_{[n]\backslash i_2}), (x_{[n]\backslash i_{d},
x'_{[n]\backslash \bar{j}_1-1}})\}$, where
$\bar{j}_1=\max(T^R\cap[i_{1}+2,i_{2}])$ when
$T^R\cap[i_{1}+2,i_{2}]\neq\emptyset$.

\begin{rem}\label{rem-LMD2-to-OMG2}
For each $X\in\{L,R\}$ and $i\in\{1,2,\cdots,6\}$, let
$$\Omega^X_{2,i}=\Omega^X_{2,i}(\bm x,\bm x')
\triangleq\bigcup_{(\bm z,\bm z')\in\Lambda^X_{2,i}}
B^{\text{S}}_{1}(\bm z,\bm z').$$ Then we have
$\Omega^X_{2}=\bigcup_{i=1}^6\Omega^X_{2,i}.$ Moreover, we can
easily obtain $\Omega^X_{2,i}$ from $\Lambda^X_{2,i}$ by Remark
\ref{rem-phi-dH2}. As an example, consider $\Omega^L_{2,4}$ with
the assumption of $|T^L\cap [i_1+1,i_{d-1}]|=0$. By Claim 2.4, we
have $\Lambda^L_{2,4}\subseteq\{(x_{[n]\backslash
k_1-1},x'_{[n]\backslash i_{d-1}}), (x_{[n]\backslash
i_1},x'_{[n]\backslash j_1'})\}$, where
$j_1'=\min(T^L\cap[i_{d-1}+1,i_{d}-1])$ when
$T^L\cap[i_{d-1}+1,i_{d}-1]\neq\emptyset$. Then
$\Omega^L_{2,4}=B^{\text{S}}_{1}(x_{[n]\backslash
k_1-1},x'_{[n]\backslash i_{d-1}})\cup
B^{\text{S}}_{1}(x_{[n]\backslash i_1},x'_{[n]\backslash j_1'})$
and by Remark \ref{rem-phi-dH2}, we have
$B^{\text{S}}_{1}(x_{[n]\backslash k_1-1},x'_{[n]\backslash
i_{d-1}})=\{\phi^{k_1-1}_{k_1;x'_{k_1-1}}(\bm x),
\phi^{k_1-1}_{i_d;x'_{i_d}}(\bm
x)\}=\{\phi^{i_{d-1}}_{k_1-1;x_{k_1}}(\bm x'),
\phi^{i_{d-1}}_{i_d;x_{i_d}}(\bm x')\}$ and
$B^{\text{S}}_{1}(x_{[n]\backslash i_1},x'_{[n]\backslash
j_1'})=\{\phi^{i_1}_{i_1;x'_{j_1'-1}}(\bm x),
\phi^{i_1}_{i_d;x'_{i_d}}(\bm
x)\}=\{\phi^{j_{1}'}_{j_1'-1;x_{j_1'}}(\bm x'),
\phi^{j_{1}'}_{i_d;x_{i_d}}(\bm x')\}$. In particular,
$|B^{\text{S}}_{1}(\bm z,\bm z')|=2$ for each $(\bm z,\bm
z')\in\Lambda^X_{2}$ and each $X\in\{L,R\}$.
\end{rem}

In the following, we prove Claims 2.1$-$2.6. Note that
$T^R=T^R(\bm x,\bm x')=T^L(\bm x',\bm x)$ and $(\bm z,\bm
z')\in\Lambda^R_\ell(\bm x,\bm x')$ if and only if $(\bm z',\bm
z)\in\Lambda^L_\ell(\bm x',\bm x)$. So, Claims 2.1$'-$2.6$'$ can
be obtained from Claims 2.1$-$2.6 directly.

The proofs of Claims 2.1$-$2.6 are similar to the proofs of Claims
1.1$-$1.3. Let
\begin{equation}\label{defeq-ka-2}
k_c=\!\left\{\!\begin{aligned}
&\max(T^L\cap[1,i_1]\backslash\{k_1,k_2\}),
~\text{if}~|T^L\cap[1,i_1]|\geq 3;\\
&1, ~\text{otherwise}.
\end{aligned}\right.
\end{equation}
and
\begin{equation}\label{defeq-pka-2}
k'_c=\!\left\{\!\begin{aligned}
&\min(T^L\cap[i_d+1,n]\backslash\{k'_1,k'_2\})-1,
~\text{if}~|T^L\cap[i_d+1,n]|\geq 3;\\
&n, ~\text{otherwise}.
\end{aligned}\right.
\end{equation}
Here, $k_1$, $k_1'$ $k_2$ and $k_2'$ are defined by
\eqref{eq-def-k1}, \eqref{eq-def-k1p}, \eqref{eq-def-k2} and
\eqref{eq-def-k2p} respectively. Then we have
\begin{align}\label{eq-all-order}
k_c<k_2<k_1\leq i_1<i_d<k'_1<k'_2\leq k'_c.\end{align}

\begin{proof}[Proof of Claim 2.1]
Note that $\Lambda^L_{2,1}(\bm x,\bm x')\neq\emptyset$ if and only
if there exists $(j,j')$ satisfying the conditions of Case 1. The
proof is similar to Claim 1.1.

For $d=2$, it is easy to see that $(j,j')$ satisfies the
conditions of Case 1 $($that is, $|S\cap[1,j-1]|=2$ and
$|T^L\cap[j+1,j']|=|S\cap[j'+1,n]|=0)$ if and only if $i_2<j\leq
j'$ and $|T^L\cap[j+1,j']|=0$. By definition of $S$ and by Lemma
\ref{iden-eq-dH}, we can prove that $x_{[j,j']}=x'_{[j,j']}$ is
contained in a run of $x_{[i_2+1,n]}=x'_{[i_2+1,n]}$, so
$\Lambda^L_{2,1}=\{(x_{[n]\backslash j},x'_{[n]\backslash j}):
j\in[i_2+1,n]\}$ and $|\Lambda^L_{2,1}|$ equals to the number of
runs of $x_{[i_2+1,n]}$.

Suppose $d\geq 3$ and $(j,j')$ satisfy the conditions of Case 1.
By $|S\cap[1,j-1]|=2$ and $|S\cap[j'+1,n]|=0$, we have
$$i_2<j\leq i_3\leq i_d\leq j'.$$ Combining this with
$|T^L\cap[j+1,j']|=0$, we can obtain $T^L\cap [i_3+1,i_d]\subseteq
T^L\cap [j+1,j']=\emptyset$, which implies that if $|T^L\cap
[i_3+1,i_d]|\neq\emptyset$, then $\Lambda^L_{2,1}(\bm x,\bm
x')=\emptyset$.

Conversely, suppose $d\geq 3$ and $T^L\cap [i_3+1,i_d]=\emptyset$.
We need to prove $\Lambda^L_{2,1}=\{(x_{[n]\backslash
i_3},x'_{[n]\backslash i_d})\}$. Let
$j_1=\max(T^L\cap[i_2+1,i_3])$ if
$T^L\cap[i_2+1,i_3]\neq\emptyset$, and $j_1=i_2+1$ otherwise. Then
by \eqref{eq-all-order}, we have
$$i_2<j_1\leq i_3\leq i_d<k'_1.$$ Clearly, $(j,j')$ satisfies
the conditions of Case 1 $($i.e., $|S\cap[1,j-1]|=2$ and
$|T^L\cap[j+1,j']|=|S\cap[j'+1,n]|=0)$ if and only if it satisfies
\begin{align*}
i_2<j_1\leq j\leq i_3\leq i_d\leq j'<k'_1.\end{align*} Moreover,
by the definition of $j_1$ and $k'_1$, we have
$$[j_1+1,i_3]\cap T=[i_d+1,k'_1-1]\cap T=\emptyset,$$
and
$$[j_1,i_3-1]\cap S=[i_d+1,k'_1-1]\cap S=\emptyset.$$
Hence, by Lemma \ref{iden-eq-dH}, $x_{[j_1,i_3]}~($resp.
$x'_{[i_d,k'_1-1]})$ is contained in a run of $\bm x~($resp. $\bm
x')$, which implies that $\Lambda^L_{2,1}=\{(x_{[n]\backslash
j},x'_{[n]\backslash j'}): i_2<j_1\leq j\leq i_3\leq i_d\leq
j'<k'_1\}=\{(x_{[n]\backslash i_3},x'_{[n]\backslash i_d})\}$.
\end{proof}

\begin{figure}[htbp]
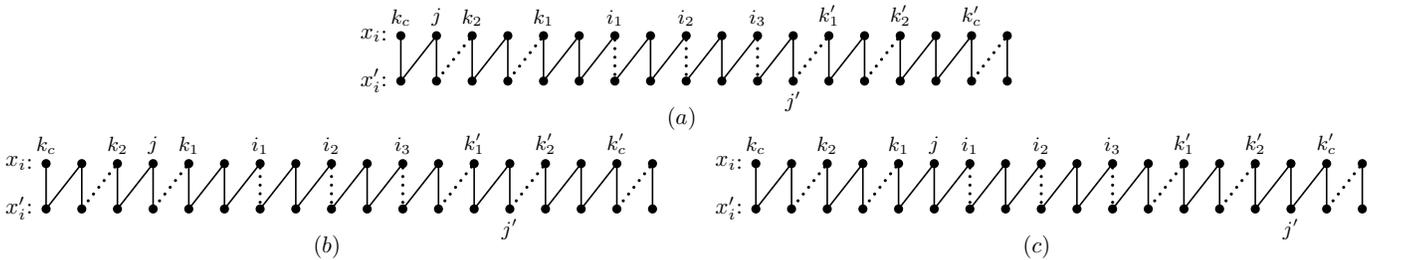

\begin{center}
\includegraphics[width=8.7cm]{cfig.7}\\
\includegraphics[width=8.7cm]{cfig.8}~~~~~~~~
\includegraphics[width=8.7cm]{cfig.9}
\end{center}
\vspace{-6pt}\caption{An illustration of the pair $(j,j')$ in Case
(iii) of the proof of Claim 2.2. Here $S=\{i_1,i_2,i_3\}$. In this
figure, (a) is for condition 1), (b) is for condition 2) and (c)
is for condition 3).}\label{fig-LMD2-2}
\end{figure}

\begin{proof}[Proof of Claim 2.2]
The proof is similar to Claim 1.2.

Suppose there exists $(j,j')$ satisfying the conditions of Case 2
$($i.e., $|T^L\cap[j+1,j']|=2$ and
$|S\cap[1,j-1]|=|S\cap[j'+1,n]|=0)$. By
$|S\cap[1,j-1]|=|S\cap[j'+1,n]|=0$, we have $$j\leq i_1<i_d\leq
j'.$$ Combining this with $|T^L\cap[j+1,j']|=2$, we have $|T^L\cap
[i_1+1,i_d]|\leq |T^L\cap[j+1,j']|=2$, which implies that
$\Lambda^L_{2,2}(\bm x,\bm x')=\emptyset$ if $|T^L\cap
[i_1+1,i_d]|\geq 3$.

Conversely, suppose $|T^L\cap [i_1+1,i_d]|\leq 2$. We have the
following Cases (i)$-$(iii).

Case (i): $|T^L\cap [i_1+1,i_d]|=2$. By \eqref{eq-all-order}, it
is easy to see that $(j,j')$ satisfying the conditions of Case 2
if and only if $k_1\leq j\leq i_1<i_d\leq j'<k'_1$. Similar to
Claim 1.1, we can prove $($by Lemma \ref{iden-eq-dH}$)$ that
$x_{[k_1,i_1]}~($resp. $x'_{[i_d,k'_1-1]})$ is contained in a run
of $\bm x~($resp. $\bm x')$, which implies that
$\Lambda^L_{2,2}=\{(x_{[n]\backslash j},x'_{[n]\backslash j'}):
k_1\leq j\leq i_1<i_d\leq j'<k'_1\}=\{(x_{[n]\backslash i_1},
x'_{[n]\backslash i_d})\}$.

Case (ii): $|T^L\cap [i_1+1,i_d]|=1$. By \eqref{eq-all-order}, it
is easy to see that $(j,j')$ satisfying the conditions of Case 2
if and only if it satisfies one of the following two conditions:
1) $k_2\leq j<k_1\leq i_1<i_d\leq j'<k'_1$; and 2) $k_1\leq j\leq
i_1<i_3<k'_1\leq j'<k'_2$. Hence, we have
$\Lambda^L_{2,2}\subseteq\{(x_{[n]\backslash j},x'_{[n]\backslash
j'}): \text{condition i) holds}\}\cup\{(x_{[n]\backslash
j},x'_{[n]\backslash j'}): \text{condition ii)
holds}\}=\{(x_{[n]\backslash k_1-1},x'_{[n]\backslash i_d}),
(x_{[n]\backslash i_1},x'_{[n]\backslash k'_1})\}$, where the
equality is obtained by applying Lemma \ref{iden-eq-dH}.

Case (iii): $|T^L\cap [i_1+1,i_d]|=0$. By \eqref{eq-all-order}, it
is easy to see that $(j,j')$ satisfying the conditions of Case 2
if and only if it satisfies one of the following three conditions:
1) $k_c\leq j<k_2$ and $i_d\leq j'<k'_1$ (see Fig.
\ref{fig-LMD2-2} (a)), which implies $(x_{[n]\backslash
k_2-1},x'_{[n]\backslash i_d})\in\Lambda^L_{2,2}$; 2) $k_2\leq
j<k_1$ and $k'_1\leq j'<k'_2$ (see Fig. \ref{fig-LMD2-2} (b)),
which implies $(x_{[n]\backslash k_1-1},x'_{[n]\backslash
k'_1})\in\Lambda^L_{2,2}$; and 3) $k_1\leq j\leq i_1$ and
$k'_2\leq j'\leq k'_c$ (see Fig. \ref{fig-LMD2-2} (c)), which
implies $(x_{[n]\backslash i_1},x'_{[n]\backslash
k'_2})\in\Lambda^L_{2,2}$. Hence, we have
$\Lambda^L_{2,2}\subseteq\big\{(x_{[n]\backslash
k_2-1},x'_{[n]\backslash i_d})$, $(x_{[n]\backslash
k_1-1},x'_{[n]\backslash k'_1})$, $(x_{[n]\backslash
i_1},x'_{[n]\backslash k'_2})\big\}$.
\end{proof}

\begin{proof}[Proof of Claim 2.3]
The proof is similar to Claim 2.1.

For $d=2$, $(j,j')$ satisfies the conditions of Case 3 $($i.e.,
$|S\cap[j'+1,n]|=2$ and $|S\cap[1,j-1]|=|T^L\cap[j+1,j']|=0)$ if
and only if $j\leq j'<i_1$ and $|T^L\cap[j+1,j']|=0$. By Lemma
\ref{iden-eq-dH}, $x_{[j,j']}=x'_{[j,j']}$ is contained in a run
of $x_{[1,i_1-1]}=x_{[1,i_1-1]}$, so
$\Lambda^L_{2,3}=\{(x_{[n]\backslash j},x'_{[n]\backslash j}):
j\in[1,i_1-1]\}$ and $|\Lambda^L_{2,3}|$ equals to the number of
runs of $x_{[1,i_1-1]}$.

Suppose $d\geq3$ and $(j,j')$ satisfies the conditions of Case 3
$($i.e., $|S\cap[j'+1,n]|=2$ and
$|S\cap[1,j-1]|=|T^L\cap[j+1,j']|=0)$. By the conditions
$|S\cap[j'+1,n]|=2$ and $|S\cap[1,j-1]|=|T^L\cap[j+1,j']|=0$, we
must have $k_1\leq j\leq i_1\leq i_{d-2}\leq j'<j_1'$ and $T^L\cap
[i_1+1,i_{d-2}]=\emptyset$, where
$j_1'=\min(T^L\cap[i_{d-2}+1,i_{d-1}]$ if
$T^L\cap[i_{d-2}+1,i_{d-1}]\neq\emptyset$, and $j_1'=i_{d-1}$
otherwise. So, if $T^L\cap [i_1+1,i_{d-2}]\neq\emptyset$, then
$\Lambda^L_{2,3}$. Conversely, if $T^L\cap
[i_1+1,i_{d-2}]=\emptyset$, then $(j,j')$ satisfies the conditions
of Case 3 if and only if $k_1\leq j\leq i_1\leq i_{d-2}\leq
j'<j_1'$. Moreover, by Lemma \ref{iden-eq-dH},
$x_{[k_1,i_1]}~($resp $x'_{[i_{d-2},j'_1-1]})$ is contained in a
run of $\bm x~($resp. $\bm x')$. Thus, we have
$\Lambda^L_{2,3}(\bm x,\bm x')=\{(x_{[n]\backslash
j},x'_{[n]\backslash j'}): k_1\leq j\leq i_1\leq i_{d-2}\leq
j'<j_1'\}=\{(x_{[n]\backslash i_1},x'_{[n]\backslash i_{d-2}})\}$.
\end{proof}

\begin{proof}[Proof of Claim 2.4]
Suppose there exists $(j,j')$ satisfying the conditions of Case 4
$($i.e., $|S\cap[1,j-1]|=0$ and
$|T^L\cap[j+1,j']|=|S\cap[j'+1,n]|=1)$. From the conditions
$|S\cap[1,j-1]|=0$ and $|S\cap[j'+1,n]|=1$, we have $$j\leq
i_1<i_{d-1}\leq j'<i_{d}.$$ Combining this with
$|T^L\cap[j+1,j']|=1$, we have $|T^L\cap [i_1+1,i_{d-1}]|\leq
|T^L\cap [j+1,j']|=1$, which implies that if $|T^L\cap
[i_1+1,i_{d-1}]|\geq 2$, then $\Lambda^L_{2,4}=\emptyset$.

Suppose $|T^L\cap [i_1+1,i_{d-1}]|=1$. $($Note that this condition
holds only if $i_1<i_{d-1}$, i.e., $d\geq 3.)$ Let
$j_1'=\min(T^L\cap[i_{d-1}+1,i_d])$ if
$T^L\cap[i_{d-1}+1,i_d]\neq\emptyset$, and $j_1'=i_d$ otherwise.
Then $(j,j')$ satisfies the conditions of Case 4 if and only if
\begin{align*}
k_1\leq j\leq i_1<i_{d-1}\leq j'<j_1'.\end{align*} Moreover, by
Lemma \ref{iden-eq-dH}, we can prove that $x_{[k_1,i_1]}~($resp
$x'_{[i_{d-1},j'_1-1]})$ is contained in a run of $\bm x~($resp.
$\bm x')$. Thus, we have $\Lambda^L_{2,4}(\bm x,\bm
x')=\{(x_{[n]\backslash j},x'_{[n]\backslash j'}): k_1\leq j\leq
i_1<i_{d-1}\leq j'<j_1'\}=\{(x_{[n]\backslash
i_1},x'_{[n]\backslash i_{d-1}})\}$.

Suppose $|T^L\cap [i_1+1,i_{d-1}]|=0$. We are to prove
$\Lambda^L_{2,4}\subseteq\{(x_{[n]\backslash
k_1-1},x'_{[n]\backslash i_{d-1}}), (x_{[n]\backslash
i_1},x'_{[n]\backslash j_1'})\}$, where
$j_1'=\min(T^L\cap[i_{d-1}+1,i_{d}-1])$ when
$T^L\cap[i_{d-1}+1,i_{d}-1]\neq\emptyset$. We need to consider the
following Cases (i) and (ii).

Case (i): $T^L\cap[i_{d-1}+1,i_d-1]\neq\emptyset$. Let
$j_2'=\min(T^L\cap[i_{d-1}+1,i_d]\backslash j_1')$ if
$T^L\cap[i_{d-1}+1,i_d]\neq\emptyset$, and $j_2'=i_d$ otherwise.
Then $(j,j')$ satisfies the conditions of Case 4 if and only if
one of the following two conditions holds: 1) $k_2\leq j<k_1\leq
i_1\leq i_{d-1}\leq j'<j_1'$; 2) $k_1\leq j\leq i_1\leq
i_{d-1}<j_1'\leq j'<j_2'$. Hence, we have
$\Lambda^L_{2,4}\subseteq\{(x_{[n]\backslash j},x'_{[n]\backslash
j'}): \text{condition 1) holds}\}\cup\{(x_{[n]\backslash
j},x'_{[n]\backslash j'}): \text{condition 2)
holds}\}=\{(x_{[n]\backslash k_1-1},x'_{[n]\backslash i_{d-1}}),
(x_{[n]\backslash i_1},x'_{[n]\backslash j'_1})\}$, where the
equality is obtained by applying Lemma \ref{iden-eq-dH}.

Case (ii): $T^L\cap[i_{d-1}+1,i_d-1]=\emptyset$. It is easy to see
that $(j,j')$ satisfies the conditions of Case 4 if and only if
$k_2\leq j<k_1\leq i_1\leq i_{d-1}\leq j'<j_1'$. Hence, we have
$\Lambda^L_{2,4}=\{(x_{[n]\backslash j},x'_{[n]\backslash j'}):
k_2\leq j<k_1\leq i_1\leq i_{d-1}\leq
j'<j_1'\}=\{(x_{[n]\backslash k_1-1},x'_{[n]\backslash
i_{d-1}})\}$.
\end{proof}

\begin{proof}[Proof of Claim 2.5]
For $d=2$, $(j,j')$ satisfies the conditions of Case 5 $($i.e.,
$|S\cap[1,j-1]|=|S\cap[j'+1,n]|=1$ and $|T^L\cap [j+1,j']|=0)$ if
and only if $i_1<j\leq j'<i_2$ and $|T^L\cap[j+1,j']|=0$. Similar
to Claim 2.1, $x_{[j,j']}=x'_{[j,j']}$ is contained in a run of
$x_{[i_1+1,i_2-1]}=x'_{[i_1+1,i_2-1]}$, so
$\Lambda^L_{2,5}=\{(x_{[n]\backslash j},x'_{[n]\backslash j}):
j\in[i_1+1,i_2-1]\}$ and $|\Lambda^L_{2,5}|$ equals to the number
of runs of $x_{[i_1+1,i_2-1]}$.

Suppose $d\geq 3$ and $(j,j')$ satisfies the conditions of Case 5.
From $|S\cap[1,j-1]|=|S\cap[j'+1,n]|=1$, we have $$i_1<j\leq
i_2\leq i_{d-1}\leq j'<i_{d}.$$ So, $T^L\cap
[i_2+1,i_{d-1}]\subseteq T^L\cap [j+1,j']=\emptyset$ (the equality
is a condition of Case 5), which implies that if $T^L\cap
[i_2+1,i_{d-1}]\neq\emptyset$, then $\Lambda^L_{2,5}=\emptyset$.

Now, suppose $d\geq 3$ and $|T^L\cap [i_2+1,i_{d-1}]|=0$. We are
to prove $\Lambda^L_{2,5}=\{(x_{[n]\backslash
i_2},x'_{[n]\backslash i_{d-1}})\}$. Let
$j_1=\max(T^L\cap[i_{1}+1,i_2])$ if
$T^L\cap[i_{1}+1,i_2\neq\emptyset$, and $j_1=i_1+1$ otherwise; let
$j_1'=\min(T^L\cap[i_{d-1}+1,i_d])-1$ if
$T^L\cap[i_{d-1}+1,i_d]\neq\emptyset$, and $j_1'=i_d$ otherwise.
Then $(j,j')$ satisfies the conditions of Case 5 if and only if it
satisfies
\begin{align*}
j_1\leq j\leq i_2\leq i_{d-1}\leq j'<j_1'.\end{align*} Moreover,
by Lemma \ref{iden-eq-dH}, we can prove that $x_{[j_1,i_2]}~($resp
$x'_{[i_{d-1},j'_1-1]})$ is contained in a run of $\bm x~($resp.
$\bm x')$. Thus, we have $\Lambda^L_{2,5}(\bm x,\bm
x')=\{(x_{[n]\backslash j},x'_{[n]\backslash j'}): j_1\leq j\leq
i_2\leq i_{d-1}\leq j'<j_1'\}=\{(x_{[n]\backslash
i_2},x'_{[n]\backslash i_{d-1}})\}$.
\end{proof}

\begin{proof}[Proof of Claim 2.6]
The proof is similar to Claim 2.4.

Suppose $(j,j')$ satisfies the conditions of Case 6 $($i.e.,
$|S\cap[j'+1,n]|=0$ and $|S\cap[1,j-1]|=|T^L\cap[j+1,j']|=1)$.
From the conditions $|S\cap[1,j-1]|=1$ and $|S\cap[j'+1,n]|=0$, we
have $$i_1<j\leq i_2<i_{d}\leq j'.$$ Combining this with
$|T^L\cap[j+1,j']|=1$, we have $|T^L\cap [i_2+1,i_{d}]|\leq
|T^L\cap [j+1,j']|=1$, which implies that if $|T^L\cap
[i_2+1,i_{d}]|\geq 2$, then $\Lambda^L_{2,6}=\emptyset$.

Suppose $|T^L\cap [i_2+1,i_{d}]|=1$. $($Note that this condition
holds only if $i_2<i_{d}$, i.e., $d\geq 3.)$ Let
$j_1=\min(T^L\cap[i_{1}+1,i_2])$ if
$T^L\cap[i_{1}+1,i_2]\neq\emptyset$, and $j_1=i_d$ otherwise. Then
$(j,j')$ satisfies the conditions of Case 6 if and only if
\begin{align*}
j_1\leq j\leq i_2<i_{d}\leq j'<k_1'.\end{align*} Moreover, by
Lemma \ref{iden-eq-dH}, we can prove that $x_{[j_1,i_2]}~($resp
$x'_{[i_{d},k'_1-1]})$ is contained in a run of $\bm x~($resp.
$\bm x')$. Thus, we have $\Lambda^L_{2,6}(\bm x,\bm
x')=\{(x_{[n]\backslash j},x'_{[n]\backslash j'}): j_1\leq j\leq
i_2<i_{d}\leq j'<k_1'\}=\{(x_{[n]\backslash i_2},x'_{[n]\backslash
i_{d}})\}$.

Suppose $|T^L\cap [i_2+1,i_{d}]|=0$. We are to prove
$\Lambda^L_{2,6}\subseteq\{(x_{[n]\backslash
i_2},x'_{[n]\backslash k'_{1}}), (x_{[n]\backslash
j_1-1},x'_{[n]\backslash i_{d}})\}$, where
$j_1=\max(T^L\cap[i_{1}+2,i_{2}])$ when
$T^L\cap[i_{1}+2,i_{2}]\neq\emptyset$. We need to consider the
following Cases (i) and (ii).

Case (i): $T^L\cap[i_{1}+2,i_2]\neq\emptyset$. Let
$j_2=\max(T^L\cap[i_{1}+2,i_2]\backslash j_1)$ if
$T^L\cap[i_{1}+2,i_2]\neq\emptyset$, and $j_2=i_1+1$ otherwise.
Then $(j,j')$ satisfies the conditions of Case 6 if and only if
one of the following two conditions holds: 1) $j_1\leq j\leq
i_2\leq i_{d}<k_1'\leq j'<k_2'$; 2) $j_2\leq j<j_1\leq i_2\leq
i_{d}\leq j'<k'_1$. Hence, we have
$\Lambda^L_{2,4}\subseteq\{(x_{[n]\backslash j},x'_{[n]\backslash
j'}): \text{condition 1) holds}\}\cup\{(x_{[n]\backslash
j},x'_{[n]\backslash j'}): \text{condition 2)
holds}\}=\{(x_{[n]\backslash i_2},x'_{[n]\backslash
k'_1}),(x_{[n]\backslash j_1-1},x'_{[n]\backslash i_{d}})\}$,
where the equality is obtained by applying Lemma \ref{iden-eq-dH}.

Case (ii): $T^L\cap[i_{1}+2,i_2]=\emptyset$. It is easy to see
that $(j,j')$ satisfies the conditions of Case 6 if and only if
$j_1\leq j\leq i_2\leq i_{d}<k_1'\leq j'<k_2'$. Hence, we have
$\Lambda^L_{2,4}=\{(x_{[n]\backslash j},x'_{[n]\backslash j'}):
j_1\leq j\leq i_2\leq i_{d}<k_1'\leq j'<k_2'\}=\{(x_{[n]\backslash
i_2},x'_{[n]\backslash k'_{1}})\}$.
\end{proof}


\section{Proof of Theorem \ref{thm-ins-size}}

In this section, we prove Theorem \ref{thm-ins-size}. Note that
from Claims 1.1$-$1.3, Claims 1.1$'-$1.3$'$, Claims 2.1$-$2.6 and
Claims 2.1$'-$2.6$'$, we can directly obtain
\begin{align*}\left|B^{\text{D,S}}_{1,1}(\bm x,\bm
x')\right|&=\left|\bigcup_{\ell=0}^2(\Omega^L_{\ell}
\cup\Omega^R_{\ell})\right|\\&\leq
\sum_{\ell=0}^2(|\Omega^L_{\ell}|+|\Omega^R_{\ell}|)\\
&\leq 2(1+(q-1)(n-1))+2(4q)+2(2(n+7))\\
&=2(q+1)n+6q+32.\end{align*} However, this bound is not tight. To
obtain a tight bound of $\big|B^{\text{D,S}}_{1,1}(\bm x,\bm
x')\big|$, we need to exclude the intersection of these subsets of
$B^{\text{D,S}}_{1,1}(\bm x,\bm x')$.

The following remark will be used to exclude repeat count of some
sequences in $\bigcup_{\ell=0}^2(\Omega^L_{\ell}
\cup\Omega^R_{\ell})$.
\begin{rem}\label{rem-LMD0-inc}
If $|T^L\cap[i_1+1,i_d]|=0$, then for any $(\bm z,\bm
z')\in\Lambda$ such that $\bm z=x_{[n]\backslash i_1}$ or $\bm
z'=x'_{[n]\backslash i_d}$, we have $B^{\text{S}}_{1}(\bm z,\bm
z')\subseteq\Omega^L_0$; if $|T^R\cap[i_1+1,i_d]|=0$, then for any
$(\bm z,\bm z')\in\Lambda$ such that $\bm z=x_{[n]\backslash i_d}$
or $\bm z'=x'_{[n]\backslash i_1}$, we have $B^{\text{S}}_{1}(\bm
z,\bm z')\subseteq\Omega^R_0$. In fact, by Claim 0, we have
$x_{[n]\backslash i_1}=x'_{[n]\backslash i_d}$ and
$\Omega^L_0=B^{\text{S}}_{1}(x_{[n]\backslash i_1})=
B^{\text{S}}_{1}(x'_{[n]\backslash i_d})$, so if $\bm
z=x_{[n]\backslash i_1}$ or $\bm z'=x'_{[n]\backslash i_d}$, then
$B^{\text{S}}_{1}(\bm z,\bm z')\subseteq
\big(B^{\text{S}}_{1}(x_{[n]\backslash i_1})\cup
B^{\text{S}}_{1}(x'_{[n]\backslash i_d})\big)=\Omega^L_0$. The
other statement can be proved similarly.
\end{rem}

To prove the upper bound of $B^{\text{D,S}}_{1,1}(\bm x,\bm x')$
in Theorem \ref{thm-ins-size}, we need the following Lemmas
\ref{lem-LMD1}$-$\ref{lem-LMD2-dH3}.
\begin{lem}\label{lem-LMD1}
For each $X\in\{L,R\}$, the following hold.
\begin{itemize}
 \item[1)] If $|T^X\cap[i_1+1,i_d]|=0$,
 then we have $\Omega^X_{1}\subseteq\Omega^X_{0}$.
 \item[2)] If $|T^X\cap[i_1+1,i_d]|\neq 0$ and $d=2$, then
 $|\Lambda^X_{1}|\leq 3$.
 \item[3)] If $|T^X\cap[i_1+1,i_d]|\neq 0$ and $d\geq 3$, then
 $|\Lambda^X_{1}|\leq 2$.
\end{itemize}
\end{lem}
\begin{proof}
We only consider $X=L$. The proof for $X=R$ is similar. By
checking Claims 1.1$-$1.3, we have the following:

1) If $|T^L\cap[i_1+1,i_d]|=0$, then for each $(\bm z,\bm
z')\in\Lambda_1$, either $\bm z=x_{[n]\backslash i_1}$ or $\bm
z'=x'_{[n]\backslash i_d}$. Hence, by Remark \ref{rem-LMD0-inc},
we have $\Omega^L_{1}=\bigcup_{(\bm z,\bm z')\in\Lambda^L_1}
B^{\text{S}}_{1}(\bm z,\bm z')\subseteq \Omega^L_{0}.$

2) If $|T^L\cap[i_1+1,i_d]|\neq 0$ and $d=2$, we have
$|\Lambda^L_{1,1}|=|\Lambda^L_{1,3}|=1$ and $|\Lambda^L_{1,2}|\leq
1$. Hence, we have $|\Lambda^L_{1}|\leq 3$.

3) If $|T^L\cap[i_1+1,i_d]|\neq 0$ and $d\geq 3$, we have $T^L\cap
[i_2+1,i_d]\neq\emptyset$ or $T^L\cap
[i_1+1,i_{d-1}]\neq\emptyset$. Then by Claims 1.1 and 1.3,
$|\Lambda^L_{1,1}|+|\Lambda^L_{1,3}|\leq 1$. Moreover, by Claim
1.2, $|\Lambda^L_{1,2}|\leq 1$. Thus, we have
$|\Lambda^L_{1}|=|\bigcup_{i=1}^3\Lambda^L_{1,i}|\leq 2$.
\end{proof}

For $d=2$, to simplify the expressions, we introduce the following
notations. For $X\in\{L,R\}$, let
$$\Lambda^X_{2,O}\triangleq\bigcup_{i\in\{1,3,5\}}\Lambda^X_{2,i}$$
and
$$\Lambda^X_{2,E}\triangleq\bigcup_{i\in\{2,4,6\}}\Lambda^X_{2,i}.$$
Correspondingly, for $X\in\{L,R\}$ and $Y\in\{O,E\}$, let
$$\Omega^X_{2,Y}\triangleq\bigcup_{(\bm z,\bm
z')\in\Lambda^X_{2,Y}}B^{\text{S}}_{1}(\bm z,\bm z')$$ and
$$\Omega_{2,Y}\triangleq\Omega^L_{2,Y}\cup\Omega^R_{2,Y}.$$
\begin{lem}\label{lem-LMD2-dH2}
Suppose $d=2$. The following hold.
\begin{itemize}
 \item[1)] $\left|\Lambda^L_{2,O}\cup
 \Lambda^R_{2,O}\right|\leq n-2$.
 \item[2)] For each $X\in\{L,R\}$, if $|T^X\cap[i_1+1,i_2]|\neq 0$,
 then $|\Lambda^X_{2,E}|\leq 6$.
 \item[3)] For each $X\in\{L,R\}$, if $|T^X\cap[i_1+1,i_2]|=0$,
 then $|\Omega^X_{2,E}\backslash\Omega^X_{0}|\leq 6$.
 \item[4)] If $|T^L\cap[i_1+1,i_2]|=|T^R\cap[i_1+1,i_2]|=0$,
 then we have $|\Omega_{0}|=2(1+(q-1)(n-1))-q=2(q-1)n-3q+2$
 and
 $|\Omega_{2}\backslash\Omega_{0}|\leq 2n-6-\delta_{q,2}$.
\end{itemize}
\end{lem}
\begin{proof}
1) Since $d=2$, then from Claims 2.1, 2.3, 2.5 and Claims 2.1$'$,
2.3$'$, 2.5$'$, we can obtain $\Lambda^L_{2,O}
=\Lambda^R_{2,O}=\big\{(x_{[n]\backslash j},x'_{[n]\backslash j}):
j\in[n]\backslash\{i_1,i_2\}\big\}.$ Hence,
$$|\Lambda^L_{2,O}\cup\Lambda^R_{2,O}|=r_1+r_2+r_3\leq n-2$$
where $r_1$
is the number of runs of $x_{[1,i_1-1]}$, $r_2$ is the number of
runs of $x_{[i_2+1,n]}$ and $r_3$ is the number of runs of
$x_{[i_1+1,i_2-1]}$.

2) Note that for $d=2$, we have
$[i_1+1,i_{d-1}]=[i_2+1,i_d]=\emptyset$. Then this statement can
be obtained directly from Claims 2.2, 2.4, 2.6 and Claims 2.2$'$,
2.4$'$, 2.6$'$.

3) As the proof for $X=R$ and for $X=L$ are similar, we only prove
the result for $X=L$. Denote $(\bm z_1,\bm
z_1')\triangleq(x_{[n]\backslash k_1-1},x'_{[n]\backslash k'_1})$,
$(\bm z_2,\bm z_2')\triangleq(x_{[n]\backslash
k_1-1},x'_{[n]\backslash i_{1}})$ and $(\bm z_3,\bm
z_3')\triangleq(x_{[n]\backslash i_2},x'_{[n]\backslash k'_1})$.
Note that $d=2$. By checking Claims 2.2, 2.4 and 2.6, we can find
that for each $(\bm z,\bm z')\in\Lambda^L_{2,E}\backslash\{(\bm
z_i,\bm z_i'): i\in\{1,2,3\}\}$, either $\bm z=x_{[n]\backslash
i_1}$ or $\bm z'=x'_{[n]\backslash i_2}$, so by Remark
\ref{rem-LMD0-inc}, we have $B^{\text{S}}_{1}(\bm z,\bm
z')\subseteq\Omega^L_{0}$. Therefore, we can obtain
$\Omega^L_{2,E} \backslash\Omega^L_{0}\subseteq
\bigcup_{i=1}^3B^{\text{S}}_{1}(\bm z_i,\bm
z_i')\subseteq\Lambda^L_{2}$, and so
$|\Omega^L_{2,E}\backslash\Omega^L_{0}|\leq
\sum_{i=1}^3|B^{\text{S}}_{1}(\bm z_i,\bm z_i')|\leq 6$.

4) Since $d=2$ and $|T^L\cap[i_1+1,i_2]|=|T^R\cap[i_1+1,i_2]|=0$,
we have $i_2=i_1+1$. $($In fact, if $i_2>i_1+1$, then we can
obtain $x_{i_1}=x'_{i_1+1}=x_{i_1+1}=x'_{i_1}$, which contradicts
to the definition of $S.)$ So, $\bm x$ and $\bm x'$ are of the
form
\begin{align*}\bm x&=\bm uab\bm v\\~\bm x'&=\bm uba\bm
v\end{align*} where $a=x_{i_1}\neq b=x_{i_2}$, $\bm
u\in\Sigma_q^{i_1-1}$ and $\bm v\in\Sigma_q^{n-i_2}$.

By Claims 0 and 0$'$, we have $\Lambda_{0}=\big\{(x_{[n]\backslash
i_1}, x'_{[n]\backslash i_2})$, $(x_{[n]\backslash i_2},
x'_{[n]\backslash i_1})\big\}$, so we can obtain
$\Omega_{0}=B^{\text{S}}_{1}(x_{[n]\backslash i_1})\cup
B^{\text{S}}_{1}(x_{[n]\backslash i_2})$. Note that
$d_{\text{H}}(x_{[n]\backslash i_1}, x_{[n]\backslash i_2})=1$.
Then $|\Omega_{0}|=|B^{\text{S}}_{1}(x_{[n]\backslash i_1})|+
|B^{\text{S}}_{1}(x_{[n]\backslash
i_2})-|B^{\text{S}}_{1}(x_{[n]\backslash i_1})\cap
B^{\text{S}}_{1}(x_{[n]\backslash i_2})|=
2(1+(q-1)(n-1))-q=2(q-1)n-3q+2$.

In the following, we prove that
$|\Omega_{2}\backslash\Omega_{0}|\leq 2n-6-\delta_{q,2}$.

By Claims 2.2, 2.4, 2.6 and Claims 2.2$'$, 2.4$'$, 2.6$'$, we
find:
\begin{itemize}
 \item
 $\Lambda^L_{2,2}=\big\{(x_{[n]\backslash k_2-1},x'_{[n]\backslash
 i_2}),(x_{[n]\backslash k_1-1},x'_{[n]\backslash k'_1})$,
 $(x_{[n]\backslash i_1},x'_{[n]\backslash k'_2})\big\}$ and
 $\Lambda^R_{2,2}=\big\{(x_{[n]\backslash i_2},x'_{[n]\backslash
 m_2-1})$, $(x_{[n]\backslash  m'_1},x'_{[n]\backslash
 m_1-1}),(x_{[n]\backslash  m'_2},x'_{[n]\backslash
 i_1})\big\}$.
 \item $\Lambda^L_{2,4}=\big\{(x_{[n]\backslash
 k_1-1},x'_{[n]\backslash i_{1}})\big\}$ and
 $\Lambda^R_{2,4}=\big\{(x_{[n]\backslash i_{1}},x'_{[n]\backslash
 m_1-1})\big\}$. Note that $d=2$ and $i_1=i_2-1$,
 so $T^L\cap[i_{d-1}+1,i_{d}-1]
 =\emptyset$ and $j_1'$ does not exists.
 \item
 $\Lambda^L_{2,6}=\big\{(x_{[n]\backslash i_2},x'_{[n]\backslash
 k'_{1}})\big\}$ and $\Lambda^R_{2,6}=\big\{(x_{[n]\backslash
 m'_{1}},x'_{[n]\backslash i_2})\big\}$.
 Note that $i_1=i_2-1$, so $T^L\cap[i_{1}+2,i_2]
 =\emptyset$ and $j_1$ does not exists.
\end{itemize}
Here $k_1$, $k'_1$, $m_1$, $m_1'$ are defined by
\eqref{eq-def-k1}$-$\eqref{eq-def-bk1p}, and $k_2$, $k'_2$, $m_2$,
$m_2'$ are defined by \eqref{eq-def-k2}$-$\eqref{eq-def-bk2p},
respectively.

For each $(\bm z,\bm z')\in\Lambda^L_{2,E}\cup\Lambda^R_{2,E}$
except for $(\bm u_1, \bm u_1')\triangleq(x_{[n]\backslash
k_1-1},x'_{[n]\backslash k'_1})$ and $(\bm u_2, \bm
u_2')\triangleq(x_{[n]\backslash m'_1},x'_{[n]\backslash m_1-1})$,
we find that either $\bm z\in\{x_{[n]\backslash
i_1},x_{[n]\backslash i_2},x'_{[n]\backslash
i_1},x'_{[n]\backslash i_2}\}$ or $\bm z'\in\{x_{[n]\backslash
i_1},x_{[n]\backslash i_2},x'_{[n]\backslash
i_1},x'_{[n]\backslash i_2}\}$, so by Remark \ref{rem-LMD0-inc},
we have $B^{\text{S}}_{1}(\bm z,\bm z')\subseteq \Omega_0$, which
implies that
$$(\Omega^L_{2,E}\cup\Omega^R_{2,E})\backslash\Omega_{0}
\subseteq\bigcup_{i=1}^2B^{\text{S}}_{1}(\bm u_i,\bm u'_i).$$

For $(\bm u_1, \bm u_1')=(x_{[n]\backslash
k_1-1},x'_{[n]\backslash k'_1})$, we have
\begin{align*}
B^{\text{S}}_{1}(x_{[n]\backslash k_1-1},x'_{[n]\backslash
k'_1})=\left\{\phi^{k_1-1}_{k_1;x'_{k_1-1}}(\bm x),
\phi^{k_1-1}_{k'_1;x'_{k'_1-1}}(\bm x)\right\}.
\end{align*} By the definition of $k_1$, we can find
$x_{k_1-1}=x'_{k_1-1}~($because $k_1-1<i_1)$, and so
$$\phi^{k_1-1}_{k_1;x'_{k_1-1}}(\bm
x)=x_{[n]\backslash k_1}.$$ We can also find
$x_{i}=x'_{i}=x_{i+1}$ for any $i\in[k_1,i_1-1]$, so
$$\phi^{k_1-1}_{k_1;x'_{k_1-1}}(\bm
x)=x_{[n]\backslash k_1}=x_{[n]\backslash i_1}\in\Omega_0.$$
Similarly, for $(\bm u_2, \bm u_2')=(x_{[n]\backslash
m'_1},x'_{[n]\backslash m_1-1})$, we have
\begin{align*}
B^{\text{S}}_{1}(\bm u_2, \bm
u_2')=\left\{\phi^{m'_1}_{m'_1-1;x'_{m'_1}}(\bm x),
\phi^{m'_1}_{m_1-1;x'_{m_1}}(\bm x)\right\},
\end{align*} and by the definition
of $m_1'$, we can find $$\phi^{m'_1}_{m'_1-1;x'_{m'_1}}(\bm
x)=x_{[n]\backslash m'_1-1}=x_{[n]\backslash i_2}\in\Omega_0.$$
Thus, we have
$$\Omega_{2,E}\backslash\Omega_{0}
\subseteq\Big\{\phi^{k_1-1}_{k'_1;x'_{k'_1-1}}(\bm
x),\phi^{m'_1}_{m_1-1;x'_{m_1}}(\bm x)\Big\}.$$

Consider $\Lambda_{2,O}\triangleq
\bigcup_{i\in\{1,3,5\}}(\Lambda^L_{2,i}\cup\Lambda^R_{2,i})$. By
Claims 2.1, 2.3, 2.5 and Claims 2.1$'$, 2.3$'$, 2.5$'$, we can
obtain:
\begin{itemize}
 \item $\Lambda^L_{2,1}=\Lambda^R_{2,1}=\big\{(x_{[n]\backslash
 j},x'_{[n]\backslash j}): j\in[i_2+1,n]\big\}$.
 \item $\Lambda^L_{2,3}=\Lambda^R_{2,3}=\big\{(x_{[n]\backslash
 j},x'_{[n]\backslash j}): j\in[1,i_1-1]\big\}$.
 \item $\Lambda^L_{2,5}=\Lambda^R_{2,5}=\emptyset$.
\end{itemize}
So, we can obtain
\begin{align*}
\Omega_{2,O}&=\big\{B^{\text{S}}_{1}(x_{[n]\backslash j})\cup
B^{\text{S}}_{1}(x'_{[n]\backslash j}):
j\in[n]\backslash\{i_1,i_2\}\big\}\\&
=\bigcup_{j\in[n]\backslash\{i_1,i_2\}}
\big\{\phi^{j}_{i_1;x'_{i_1}}(\bm x),\phi^{j}_{i_2;x'_{i_2}}(\bm
x)\big\},\end{align*} where the second equation holds because for
each $j\in[n]\backslash\{i_1,i_2\}$, $x_{[n]\backslash j}$ and
$x'_{[n]\backslash j}$ are of the form
\begin{align*}x_{[n]\backslash j}&=\bm u'ab\bm v'\\
~x'_{[n]\backslash j}&=\bm u'ba\bm v'\end{align*} where $\bm u'\bm
v'$ is obtained from $\bm u\bm v$ by a single deletion. For
$j=i_1-1$, we find
\begin{align*}
\phi^{i_1-1}_{i_1;x'_{i_1}}(\bm x)=\phi^{i_1}_{i_1-1;x'_{i_1}}(\bm
x)\in\Omega_0.
\end{align*} Moreover, we find
\begin{align*}\phi^{i_1-1}_{i_2;x'_{i_2}}(\bm x)&=\bm u'aa\bm v\\
~x_{[n]\backslash i_2}&=\bm u'ca\bm v\end{align*} where $\bm
u'=x_{[1,i_1-2]}$ and $c=x_{i_1-1}$. Hence, we have
$\phi^{i_1-1}_{i_2;x'_{i_2}}(\bm x)\in
B^{\text{S}}_{1}(x_{[n]\backslash i_2})\subseteq\Omega_0$, which
implies that $\{\phi^{i_1-1}_{i_1;x'_{i_1}}(\bm
x),\phi^{i_1-1}_{i_2;x'_{i_2}}(\bm x)\}\subseteq\Omega_0$.
Similarly, we can prove that $\{\phi^{i_2+1}_{i_1;x'_{i_1}}(\bm
x),\phi^{i_2+1}_{i_2;x'_{i_2}}(\bm x)\}\subseteq\Omega_0$. Thus,
\begin{align*}
\Omega_{2,O}=\bigcup_{j\in[n]\backslash[i_1-1,i_2+1]}
\{\phi^{j}_{i_1;x'_{i_1}}(\bm x),\phi^{j}_{i_2;x'_{i_2}}(\bm
x)\}\end{align*} and so
$$|\Omega_{2,E}\backslash\Omega_0|+|\Omega_{2,O}\backslash\Omega_0|
\leq 2+2(n-4)=2n-6.$$

Note that if $i_1=1$ or $i_2=n$, then $k_1$ and $m_1'$ do not
exist and so $\phi^{k_1-1}_{k'_1;x'_{k'_1-1}}(\bm x)$ and
$\phi^{m'_1}_{m_1-1;x'_{m_1}}(\bm x)$ do not exist. Thus,
$|\Omega_{2,E}\backslash\Omega_0|=0$ and
$|\Omega_{2,O}\backslash\Omega_0|\leq 2(n-3)$, and so still
$$|\Omega_{2,E}\backslash\Omega_0|+|\Omega_{2,O}\backslash\Omega_0|
\leq 2n-6.$$

If $p=2$, consider $j=i_1-2$. We find that
\begin{align*}\phi^{i_1-2}_{i_1;x'_{i_1}}(\bm x)&=\bm u''cbb\bm v\\
~x_{[n]\backslash i_1}&=\bm u''ecb\bm v\end{align*} and
\begin{align*}\phi^{i_1-2}_{i_2;x'_{i_2}}(\bm x)&=\bm u''caa\bm v\\
~x_{[n]\backslash i_2}&=\bm u''eca\bm v\end{align*} where $\bm
u''=x_{[1,i_1-3]}$, $c=x_{i_1-1}$ and $e=x_{i_1-2}$. Since $p=2$,
then either $c=a$ or $c=b$. Hence, either
$d_{\text{H}}(\phi^{i_1-2}_{i_1;x'_{i_1}}(\bm x),x_{[n]\backslash
i_1})=1$ or $d_{\text{H}}(\phi^{i_1-2}_{i_2;x'_{i_2}}(\bm
x),x_{[n]\backslash i_2})=1$, and so either
$\phi^{i_1-2}_{i_1;x'_{i_1}}(\bm x)\in
B^{\text{S}}_{1}(x_{[n]\backslash i_1})\subseteq\Omega_0$ or
$\phi^{i_1-2}_{i_2;x'_{i_2}}(\bm x)\in
B^{\text{S}}_{1}(x_{[n]\backslash i_2})\subseteq\Omega_0$, which
implies that $|\{\phi^{i_1-2}_{i_1;x'_{i_1}}(\bm
x),\phi^{i_1-2}_{i_2;x'_{i_2}}(\bm x)\}\backslash\Omega_0|=1$.
Similarly, we can prove that $|\{\phi^{i_2+2}_{i_1;x'_{i_1}}(\bm
x),\phi^{i_2+2}_{i_2;x'_{i_2}}(\bm x)\}\backslash\Omega_0|=1$. So,
$$|\Omega_{2,E}\backslash\Omega_0|+|\Omega_{2,O}\backslash\Omega_0|
\leq 2n-6-2=2n-8.$$ On the other hand, if $i_1=2~($or $i_2=n-1)$,
by $q=2$ and $x_1=x'_1~($resp. $x_n=x_n')$, we can find $x_1=x'_2$
or $x'_1=x_2~($resp. $x_n=x'_{n-1}$ or $x_{n-1}=x'_n)$, and so
$k_1$ or $m_1$ does not exists $($resp. $k'_1$ or $m'_1$ does not
exists$)$, which implies that still
$|\Omega_{2,E}\backslash\Omega_0|+|\Omega_{2,O}\backslash\Omega_0|
\leq 2n-8~($because $|\Omega_{2,E}\backslash\Omega_0|=1$ and
$|\Omega_{2,O}\backslash\Omega_0|\leq 2n-9)$. If $i_1=1$ or
$i_2=n$, then we have
$|\Omega_{2,E}\backslash\Omega_0|+|\Omega_{2,O}\backslash\Omega_0|
\leq 2n-7~($because $|\Omega_{2,E}\backslash\Omega_0|=0$ and
$|\Omega_{2,O}\backslash\Omega_0|\leq 2n-7)$.

By the above discussions, we can obtain
$|\Omega_{2}\backslash\Omega_{0}|\leq 2n-6-\delta_{q,2}$, where
$\delta_{q,2}=1$ if $q=2$, and $\delta_{q,2}=0$ otherwise.
\end{proof}

\begin{lem}\label{lem-LMD2-dH3}
Suppose $d\geq 3$. The following hold.
\begin{itemize}
 \item[1)] For each $X\in\{L,R\}$, if $|T^X\cap[i_1+1,i_d]|=0$,
 then $|\Omega^X_{2}\backslash\Omega^X_{0}|\leq 8$.
 \item[2)] For each $X\in\{L,R\}$, if $|T^X\cap[i_1+1,i_d]|\neq 0$,
 then $|\Lambda^X_{2}|\leq 8$.
\end{itemize}
\end{lem}
\begin{proof}
1) For $X=L$, the result can be obtained from Claims 2.1$-$2.6 and
Remark \ref{rem-LMD0-inc}; for $X=R$, the result can be obtained
from Claims 2.1$'-$2.6$'$ and Remark \ref{rem-LMD0-inc}.

2) The result can be obtained directly from Claims 2.1$-$2.6
$($for $X=L)$ and Claims 2.1$'-$2.6$'~($for $X=R)$.
\end{proof}

Now, we can prove Theorem \ref{thm-ins-size}.

\begin{proof}[Proof of Theorem \ref{thm-ins-size}]
We first prove that if $d=d_{\text{H}}(\bm x,\bm x')\geq 2$ and
$n\geq\max\{\frac{q+23}{2}, \frac{5q+19}{q-1}\}$, then
$$|B^{\text{D,S}}_{1,1}(\bm x,\bm x')|\leq 2qn-3q-2-\delta_{q,2}.$$
We divide our discussions into the following two cases.

Case 1: $d\geq 3$.
\begin{itemize}
\item If $|T^L\cap[i_1+1,i_d]|=|T^R\cap[i_1+1,i_d]|=0$, then by
Claim 0, Claim 0$'$, 1) of Lemma \ref{lem-LMD1}, and 1) of Lemma
\ref{lem-LMD2-dH3}, we have $|B^{\text{D,S}}_{1,1}(\bm x,\bm
x')|\leq 2(1+(q-1)(n-1))+2(8)=2(q-1)n-2q+20\leq
2qn-3q-2-\delta_{q,2}$, where the last inequality comes from the
assumption that $n\geq\max\{\frac{q+23}{2}, \frac{5q+19}{q-1}\}$.

\item If $|T^L\cap[i_1+1,i_d]|=0$ and $|T^R\cap[i_1+1,i_d]|\neq
0$, then by Claim 0, 1) of Lemma \ref{lem-LMD1}, 1) of Lemma
\ref{lem-LMD2-dH3}; 3) of Lemma \ref{lem-LMD1}, and 2) of Lemma
\ref{lem-LMD2-dH3}, we have $|B^{\text{D,S}}_{1,1}(\bm x,\bm
x')|\leq (1+(q-1)(n-1))+8+q(2)+2(8)=(q-1)n+q+26\leq
2qn-3q-2-\delta_{q,2}$.

\item If $|T^L\cap[i_1+1,i_d]|\neq 0$ and
$|T^R\cap[i_1+1,i_d]|=0$, then by Claim 0$'$, 1) of Lemma
\ref{lem-LMD1}, 1) of Lemma \ref{lem-LMD2-dH3}, 3) of Lemma
\ref{lem-LMD1}, and 2) of Lemma \ref{lem-LMD2-dH3}, we have
$|B^{\text{D,S}}_{1,1}(\bm x,\bm x')|\leq
(1+(q-1)(n-1))+8+q(2)+2(8)=(q-1)n+q+26\leq 2qn-3q-2-\delta_{q,2}$.

\item If $|T^L\cap[i_1+1,i_d]|\neq 0$ and
$|T^R\cap[i_1+1,i_d]|\neq 0$, then by Claim 0, Claim 0$'$, 3) of
Lemma \ref{lem-LMD1}, and 2) of Lemma \ref{lem-LMD2-dH3}, we have
$|B^{\text{D,S}}_{1,1}(\bm x,\bm x')|\leq 2(q(2)+2(8))=4q+32\leq
2qn-3q-2-\delta_{q,2}$.
\end{itemize}

Case 2: $d=2$.
\begin{itemize}
\item If $|T^L\cap[i_1+1,i_2]|=|T^R\cap[i_1+1,i_2]|=0$, then by
Claim 0, Claim 0$'$, 1) of Lemma \ref{lem-LMD1} and 4) of Lemma
\ref{lem-LMD2-dH2}, we have $|B^{\text{D,S}}_{1,1}(\bm x,\bm
x')|\leq 2(1+(q-1)(n-1))-q+2n-6-\delta_{q,2}=
2qn-3q-2-\delta_{q,2}$.

\item If $|T^L\cap[i_1+1,i_2]|=0$ and $|T^R\cap[i_1+1,i_2]|\neq
0$, then by Claim 0, 1) of Lemma \ref{lem-LMD1}, 3) of Lemma
\ref{lem-LMD2-dH2}; 2) of Lemma \ref{lem-LMD1}, and 1)$-$2) of
Lemma \ref{lem-LMD2-dH2}, we have $|B^{\text{D,S}}_{1,1}(\bm x,\bm
x')|\leq (1+(q-1)(n-1))+6+q(3)+2(n-2+6)=(q+1)n+2q+16\leq
2qn-3q-2-\delta_{q,2}$, where the last inequality comes from the
assumption that $n\geq\max\{\frac{q+23}{2}, \frac{5q+19}{q-1}\}$.

\item If $|T^L\cap[i_1+1,i_2]|\neq 0$ and
$|T^R\cap[i_1+1,i_2]|=0$, then by Claim 0$'$, 1) of Lemma
\ref{lem-LMD1}, 3) of Lemma \ref{lem-LMD2-dH2}; 2) of Lemma
\ref{lem-LMD1}, and 1)$-$2) of Lemma \ref{lem-LMD2-dH2}, we have
$|B^{\text{D,S}}_{1,1}(\bm x,\bm x')|\leq
(1+(q-1)(n-1))+6+q(3)+2(n-2+6)=(q+1)n+2q+16\leq
2qn-3q-2-\delta_{q,2}$.

\item If $|T^L\cap[i_1+1,i_D]|\neq 0$ and
$|T^R\cap[i_1+1,i_0]|\neq 0$, then by 2) of Lemma \ref{lem-LMD1},
and 1)$-$2) of Lemma \ref{lem-LMD2-dH2}, we have
$|B^{\text{D,S}}_{1,1}(\bm x,\bm x')|\leq
2q(3)+2(n-2+6+6)=2n+6q+20\leq
2qn-3q-2-\delta_{q,2}$.
\end{itemize}


Thus, we can obtain $|B^{\text{D,S}}_{1,1}(\bm x,\bm x')|\leq
2qn-3q-2-\delta_{q,2}$.

To prove the tightness of this bound, we consider the following
two examples.

\begin{exam}\label{exm-int-size-1-q3}
Let $q\geq 3$ and $n\geq 5$. Let $\bm x,\bm x'\in\Sigma_q^n$ such
that \begin{align*}\bm x&=01201A_{n-5}(01)\\
\bm x'&=10201A_{n-5}(01).\end{align*} We have
$S=\{i_1,i_2\}=\{1,2\}$, where $S$ is defined according to
\eqref{eq-def-S}. It is not hard to verify that
\begin{itemize}
 \item $\Omega_0=B^{\text{S}}_{1}(x_{[n]\backslash 1})\cup
 B^{\text{S}}_{1}(x_{[n]\backslash 2})$ and
 $d_{\text{H}}(x_{[n]\backslash 1},x'_{[n]\backslash 2})=1$, where
 $x_{[n]\backslash 1}=1201A_{n-5}(01)$ and
 $x_{[n]\backslash 2}=0201A_{n-5}(01)$, so
 $|\Omega_0|=2(1+(q-1)(n-1))-q=2(q-1)n-3q+4$.
 \item Let $\Omega_{2}'\triangleq\bigcup_{j=4}^n
 \big\{\phi^{j}_{1;1}(\bm x),\phi^{j}_{2;0}(\bm
 x)\big\}$. Then $\Omega_{2}'\subseteq\Omega_{2,O}$
 and $\big|\Omega_{2}'|=2(n-3)$. Moreover, for each
 $\bm z\in\Omega_2'$ and each
 $\bm z'\in\{x_{[n]\backslash 1}, x_{[n]\backslash 2}\}$, we have
 $d_{\text{H}}(\bm z,\bm z')\geq 2$, and so $\Omega_2'\cap\Omega_0
 =\emptyset$, which implies that
 $\Omega_2'\subseteq\Omega_{2,O}\backslash\Omega_0$.
\end{itemize} Thus, we can obtain $$|B^{\text{D,S}}_{1,1}(\bm
x,\bm x')|\geq |\Omega_0|+|\Omega_2'|=2qn-3q-2.$$
\end{exam}

\begin{exam}\label{exm-int-size-1-q2}
Let $q=2$ and $n\geq 4$. Let $\bm x,\bm x'\in\Sigma_q^n$ such
that \begin{align*}\bm x&=0101A_{n-4}(01)\\
\bm x'&=1001A_{n-4}(01).\end{align*} For this example, it is not
hard to verify that
\begin{itemize}
 \item $\Omega_0=B^{\text{S}}_{1}(x_{[n]\backslash 1})\cup
 B^{\text{S}}_{1}(x_{[n]\backslash 2})$ and
 $d_{\text{H}}(x_{[n]\backslash 1},x'_{[n]\backslash 2})=1$, so
 $|\Omega_0|=2(1+(q-1)(n-1))-q=2(q-1)n-3q+4$.
 \item Let $\Omega_2'\triangleq\bigcup_{j=4}^n
 \big\{\phi^{j}_{1;1}(\bm x),\phi^{j}_{2;0}(\bm
 x)\big\}\backslash\{\phi^{4}_{2;0}(\bm x)\}$.
 Then $\Omega_2'\subseteq\Omega_2$ and
 $\big|\Omega_2'\big|=2(n-3)-1$. Moreover,
 for each $\bm z\in\Omega_2'$ and each
 $\bm z'\in\{x_{[n]\backslash 1}, x_{[n]\backslash 2}\}$, we have
 $d_{\text{H}}(\bm z,\bm z')\geq 2$, and so
 $\Omega_2'\subseteq\Omega_{2,O}\backslash\Omega_0$.
 Note that in this example, we can see that
 $\phi^{4}_{2;0}(\bm x)=000A_{n-4}(01)$ and
 $x_{[n]\backslash 2}=001A_{n-4}(01)$, so $\phi^{4}_{2;0}(\bm x)\in
 B^{\text{S}}_{1}(x_{[n]\backslash 2})\subseteq\Omega_0$.
\end{itemize} Hence, we have $$|B^{\text{D,S}}_{1,1}(\bm
x,\bm x')|\geq |\Omega_0|+|\Omega_2'|=2qn-3q-3=4n-9.$$
\end{exam}

By the above discussions, we proved Theorem \ref{thm-ins-size}.
\end{proof}

\begin{rem}\label{rem-G2-dL2-dH3}
By the definition of $T^L$ and $T^R$, it is easy to see that if
the Levenshtein distance $d_{\text{L}}(\bm x,\bm x')\geq 2$, then
we must have $|T^L\cap[i_1+1,i_d]|\neq 0$ and
$|T^R\cap[i_1+1,i_d]|\neq 0$. Therefore, if $d_{\text{H}}(\bm
x,\bm x')\geq 3$ and $d_{\text{L}}(\bm x,\bm x')\geq 2$, then by
the proof of Theorem \ref{thm-ins-size}, we can obtain
$|B^{\text{D,S}}_{1,1}(\bm x,\bm x')|\leq 4q+32$, which depends
only on $q$. However, this bound is not tight. To obtain a tight
bound independent of $n$ for this case, more careful discussions
are needed. This problem will be investigated in our future work.
\end{rem}

\section{Conclusions and future work}

We proved a tight upper bound on the intersection size of error
balls of single-deletion single-substitution channel for any
$q$-ary sequences $\bm x, \bm x'$ of length $n$ and with Hamming
distance $d_{\text{H}}(\bm x,\bm x')\geq 2$. This upper bound is
the minimum number of channel outputs $($reads$)$ required to
reconstruct a sequence in a code with minimum Hamming distance
$2$.

The bound obtained in this work depends on the sequence length
$n$. If we consider any $\bm x, \bm x'\in\Sigma_q^n$ with Hamming
distance $d_{\text{H}}(\bm x,\bm x')\geq 3$ and Levenshtein
distance $d_{\text{L}}(\bm x,\bm x')\geq 2$, then as pointed out
in Remark \ref{rem-G2-dL2-dH3}, we can obtain an upper bound of
$|B^{\text{D,S}}_{1,1}(\bm x,\bm x')|$ depending only on $q$. For
binary code, this requirement can be satisfied by introducing a
redundancy of only $\log n$ bits. The problem of constructing
reconstruction codes with constant number of reads (i.e., the
number of reads is independent of $n$ and depend only on $q$) for
single-deletion single-substitution channel is left in our future
work.

Another interesting problem is to generalize the method to
single-deletion $s$-substitution channel, that is, to derive a
tight upper bound of $\big|B^{\text{D,S}}_{1,s}(\bm x,\bm
x')\big|$, where $s\geq 2$ is any fixed integer. We need to
consider the set $\big\{(x_{[n]\backslash j}, x'_{[n]\backslash
j'}): j,j'\in[n]~\text{and}~ d_{\text{H}}(x_{[n]\backslash j'},
x'_{[n]\backslash j})\leq 2s\big\}$ and can divide it by the
similar method of this paper. Correspondingly,
$B^{\text{D,S}}_{1,s}(\bm x,\bm x')$ can be divided into some
subsets and each subset can be easily determined. However, the
difficulty is how to find the intersection of these subsets.

\end{document}